\documentclass[12pt]{article}
\usepackage[dvips]{graphicx} 
\usepackage{latexsym}
\usepackage[authoryear]{natbib}
\let \chapter \section
\usepackage{enumerate}
\usepackage{multirow}
\usepackage{xfrac}
\usepackage{lmodern}
\usepackage{epsfig}
\usepackage{graphics,graphicx,color}
\usepackage{amssymb,amsfonts,amsthm,amsmath}
\usepackage{multirow, booktabs}
\usepackage{mathrsfs}

\newcommand{\argmax}{\operatornamewithlimits{argmax}}
\newcommand{\argmin}{\operatornamewithlimits{argmin}}

\usepackage{amsmath,bbm,amssymb,mathrsfs}
\usepackage{amsthm,amscd,float}             
\usepackage{graphicx, multirow, graphics}
\usepackage{comment}
\usepackage{longtable, lscape}
\usepackage{rotating}
\usepackage{dcolumn}
\usepackage{natbib}
\usepackage{booktabs,caption,fixltx2e}
\usepackage[flushleft]{threeparttable}
\usepackage{caption}
\usepackage{hyperref}

\usepackage{subfigure}

\newtheorem{theorem}{Theorem}
\newtheorem{lemma}{Lemma}
\newtheorem{corollary}{Corollary}

\newtheorem{example}{Example}

\newcommand{\bm}[1]{\mbox{\boldmath$ #1 $\unboldmath}}

\newcommand{\nx}[1]{{\color{black}{#1}}}

\usepackage{epstopdf}

\begin{document}
	
	\begin{center}
		{\Large \bf Sequential Design of Computer Experiments with Quantitative and Qualitative Factors in Applications to HPC Performance Optimization}\\
		\vskip 5pt
		Xia Cai$^{\dag}$, Li Xu$^{\ddag}$, C. Devon Lin$^{\S}$, Yili Hong$^{\ddag}$ and Xinwei Deng$^{\ddag}$\footnote{Address for correspondence: Xinwei Deng, Associate Professor, Department of Statistics,
			Virginia Tech, Blacksburg, VA 24061 (E-mail: xdeng@vt.edu).} \\
		\vskip 5pt
		$^{\dag}$School of Science, Hebei University of Science and Technology, China\\
		$^{\ddag}$Department of Statistics, Virginia Tech, USA\\
		$^{\S}$Department of Mathematics and Statistics, Queen's University, Canada\\
	\end{center}
	
\begin{abstract}
Computer experiments with both qualitative and quantitative factors are widely used in many applications. 
Motivated by the emerging need of optimal configuration in the high-performance computing (HPC) system, 
this work proposes a sequential design, denoted as adaptive composite exploitation and exploration (CEE), for optimization of computer experiments with qualitative and quantitative factors.
The proposed adaptive CEE method combines the predictive mean and standard deviation based on the additive Gaussian process
to achieve a meaningful balance between exploitation and exploration for optimization.
Moreover, the adaptiveness of the proposed sequential procedure allows the selection of next design point from the adaptive design region.
Theoretical justification of the adaptive design region is provided. 
The performance of the proposed method is evaluated by several numerical examples in simulations. 
The case study of HPC performance optimization further elaborates the merits of the proposed method. 
\end{abstract}

	\newpage
	
\section{Introduction}
In many areas of the fourth industrial revolution, high-performance computing (HPC) provides important infrastructures for enabling large-scale data analytics.
Reliable computing performance is vital for cloud computing, data storage and management, and optimization (Sakellariou et al., 2018).
Thus, the investigation of performance variability of HPC has drawn great attention in recent research (Cameron et al., 2019).
The variability of HPC performance exists in several aspects, of which  the input/output (IO) variability is of great interest.
The IO performance is usually measured by  the IO throughput (i.e., data transfer speed), which can vary from run to run.
The variability of IO throughput can be affected by various system factors such as CPU frequency, the number of threads, IO operation mode, and IO scheduler, through a complicated relationship (Cameron et al., 2019).

To configure an HPC system with reliable IO performance, one important task is to find an optimal configuration (i.e., a certain level combination of system factors) that optimizes the IO performance measure. 
The search for the optimized configuration is a challenging task since the functional relationship between IO performance measure and system factors is unknown and complicated, especially for the HPC system containing both quantitative and qualitative inputs. 
To address this challenge, sequential designs in computer experiments (Sacks et al. 1989; Santner et al. 2003; Fang et al. 2005) can be used. 
It is a novel application of sequential designs of experiments for the HPC performance optimization.


The execution of computer experiment of HPC is time consuming.
For example, it can take hours or days to collect the HPC IO performance in a single run under certain system configurations.
Therefore, statistical surrogates are often adopted for statistical analysis and uncertainty quantification (Sacks et al. 1989; Bingham et al. 2014).
One fundamental issue is the design of experiments, i.e., how to choose the settings of input variables to run computer experiments to obtain the output responses 
for the objectives of interest.
The commonly used designs are space-filling designs (Lin and Tang 2015; Joseph 2016; Xiao and Xu 2017; Wang et al. 2018).
To entertain both qualitative and quantitative inputs, space-filling designs such as sliced Latin hypercube designs and marginally coupled designs have been introduced (Qian 2012; Deng et al. 2015; He et al. 2017; He et al. 2019). 
However, these designs are proposed with the aim of building an accurate emulator and thus they are not designated for other objectives such as the optimization problem we consider here.
An objective-oriented design approach is to use sequential designs which  find the new input setting sequentially for the objective of interest (Picheny et al. 2016; Sauer et al. 2020). 
Such an approach has appeared being efficient and advantageous as indicated in many applications (Gramacy 2020).
For example,  Bingham et al. (2014) adopted sequential designs for choosing input settings of a computer simulator for the maximization of the tidal power in the Bay of Fundy, Nova Scotia, Canada (Ranjan et al. 2011).
One popular approach in the sequential design framework is to use an expected improvement (EI) criterion (Jones et al. 1998; Schonlau et al. 1998; S\'obester et al. 2005; Ponweiser et al. 2008). An EI criterion was initially introduced  for the global optimization of black box functions (computer simulators) by  Jones et al. (1998). 
Since then, various EI criteria have been proposed for other objectives such as contour estimation (Ranjan et al. 2008), quantile estimation (Roy 2008), estimating the probability of rare events and system failure (Bichon et al. 2009), and prediction (Yang et al. 2020). Other criteria in the sequential design framework include the upper confidence bound (Srinivas et al. 2010)), the knowledge gradient method (Frazier et al. 2008; Scott et al. 2011), and hierarchical expected improvement (Chen et al. 2019).  
However, to the best of our knowledge, these sequential design approaches including those using EI criteria have exclusively focused on computer experiments with only quantitative inputs.
These approaches may not be directly applicable to computer experiments, such as the HPC experiment, with both qualitative and quantitative factors.

In this article, our scope is to develop a sequential design approach for efficient optimization of computer experiments with both qualitative and quantitative (QQ) factors. In the HPC example,  the IO operation mode is a qualitative variable, while the CPU frequency is a quantitative variable. 
We propose an adaptive {\em composite exploitation and exploration} (CEE) method for the global optimization for computer experiments with QQ factors. 
A new criterion combining the predictive mean and standard deviation based on the additive Gaussian process (AGP) (Deng et al. 2017) is introduced to search for follow-up design points. 
Similar to the EI and other criteria, the proposed criterion also aims to  achieve the balance between exploitation and exploration when searching for the next input setting. 
What is fundamentally different and makes this criterion novel is that the search design region at each stage via the new criterion is adaptive in the sense that the design region changes with the data collected. 
Theoretical justifications are provided to support the choice of the adaptive design region. 
In addition, the proposed CEE criterion has a simple expression with meaningful interpretation to choose the next design point sequentially based on the AGP as the surrogate.
The sequential design procedure with the proposed criterion appears to be efficient in both computation and finding the optimal setting, i.e., the setting of optimizing the response output.


The remainder of this paper is organized as follows. Section 2 briefly reviews the additive Gaussian process model.
Section 3 presents the details of the proposed adaptive CEE method and its theoretical justification on the choice of adaptive design region.
In Section 4, several numerical examples are conducted to illustrate the effectiveness of the proposed method.
Section 5 presents the case study of HPC experiments, where the proposed method is demonstrated to efficiently find the optimal setting for HPC performance optimization.
We conclude this work with some discussion in Section 6.
	
	\section{Brief Review of Additive Gaussian Process Model}
	Consider a computer experiment with $p$ quantitative factors ${\bm x}=(x_1,\cdots,x_p)^T\in {\bm X}\subseteq R^p$ and $q$ qualitative factors ${\bm z}=(z_1,\cdots,z_q)^T\in{\bm Z}$ with the $j$th qualitative factor having $m_j$ levels, $j=1,\cdots,q$, and the corresponding output is denoted by $Y$, where  ${\bm Z}$ contains $M = \prod_{j=1}^{q} m_{j}$ elements.
	Suppose that the observed data are $({\bm w}_t^T, y_t), t=1,\cdots,n$, where ${\bm w}_t=({\bm x}_t^T,{\bm z}_t^T)^T=(x_{t1},\cdots,x_{tp},z_{t1},\cdots,z_{tq})^T$.
	To model the relationship between output $Y$ and input ${\bm w}$, the AGP model assumes
	\begin{align}\label{model}
		Y({\bm x},z_1,\cdots,z_q)=\mu+G_1({\bm x},z_1)+\cdots+G_q({\bm x},z_q),
	\end{align}
	where $\mu$ is the overall mean, and the $G_j$'s are independent Gaussian processes with mean zero and covariance function $\phi_j$. 
	For two inputs ${\bm w}_1=({\bm x}_1^T,{\bm z}_1^T)^T=(x_{11},\cdots,x_{1p},z_{11},\cdots,z_{1q})^T$
	and 
	${\bm w}_2=({\bm x}_2^T,{\bm z}_2^T)^T=(x_{21},\cdots,x_{2p},z_{21},\cdots,z_{2q})^T$,
	the covariance function $\phi_j$ is given by
	\begin{align}
		\phi_j(G_j({\bm x}_1,z_{1j}),G_j({\bm x}_2,z_{2j})) = \sigma_j^2\tau^{(j)}_{z_{1j},z_{2j}}R({\bm x}_1,{\bm x}_2|{\bm\theta}^{(j)}),
	\end{align}
	where $\sigma_j^2$ is the variance component associated with $G_j$, $\tau_{r,s}^{(j)}$ is the correlation of the $r$th level and the $s$th level of the qualitative factor $z_j, j=1,\cdots,q$.
	That is, $\tau_{r,s}^{(j)}$ is the $(r,s)$th element in correlation matrix ${\bm T}^{(j)}=(\tau_{r,s}^{(j)})_{m_j\times m_j}$, $j = 1, \ldots, q$.
	A common choice of the correlation function is the Gaussian correlation function $R({\bm x}_1,{\bm x}_2|{\bm\theta}^{(j)})=\exp\left\{-\sum_{i=1}^p\theta_i^{(j)}(x_{1i}-x_{2i})^2\right\}$
	for any two quantitative inputs ${\bm x}_1$ and ${\bm x}_2$, where ${\bm\theta}^{(j)}=(\theta_1^{(j)},\cdots,\theta_p^{(j)})^T$ (Deng et al. 2017). Then, the response $Y$ follows a Gaussian process with mean zero and the covariance function $\phi$ specified by
	\begin{align}
		\phi(Y({\bm w}_1),Y({\bm w}_2))&=cov(Y({\bm x}_1,{\bm z}_1),Y({\bm x}_2,{\bm z}_2))\notag\\
		&=\sum_{j=1}^q\sigma_j^2\tau^{(j)}_{z_{1j},z_{2j}}R({\bm x}_1,{\bm x}_2|{\bm\theta}^{(j)})\notag\\
		&=\sum_{j=1}^q\sigma_j^2\tau^{(j)}_{z_{1j},z_{2j}}\exp\left\{-\sum_{i=1}^p\theta_i^{(j)}(x_{1i}-x_{2i})^2\right\}.
	\end{align}
	
	We denote $Y_0=Y({\bm w}_0)$ as the prediction of $Y$ at a new setting ${\bm w}_0=({\bm x}_0^T,{\bm z}_0^T)^T$. 
	Let ${\bm y}_n=(y_1,\cdots,y_n)^T$ be $n$ outputs from the input $({\bm w}_1^T,\cdots,{\bm w}_n^T)^T$.
	Based on the AGP, it is easy to obtain that $Y_0|{\bm y}_n$ follows a normal distribution with
	\begin{align}
		E(Y_0|{\bm y}_n)   &= \mu_{0|n}=\mu+{\bm r}_0^T{\bm\Phi}^{-1}({\bm y}_n-\mu{\bm 1}_n), \label{eq: pred-mean}\\
		Var(Y_0|{\bm y}_n) &=  \sigma_{0|n}^2=\sum_{j=1}^q\sigma_j^2-{\bm r}_0^T{\bm\Phi}^{-1}{\bm r}_0, \label{eq: pred-var}
	\end{align}
	where ${\bm\Phi}$ is the covariance matrix of $\bm y_n$, and ${\bm r}_0=(\phi_{01},\cdots,\phi_{0n})^T$ with $\phi_{0t}$ given by
	$\phi_{0t}=\phi(Y({\bm w}_0),Y({\bm w}_t))=\sum_{j=1}^q\sigma_j^2\tau^{(j)}_{z_{0j},z_{tj}}\exp\left\{-\sum_{i=1}^p\theta_i^{(j)}(x_{0i}-x_{ti})^2\right\}$, $t=1,2,\ldots,n$.

	Clearly the mean and variance of $Y_0|\bm y_n$, i.e., $\mu_{0|n}$ and $\sigma_{0|n}^2$, involve the parameters $\mu$, ${\bm\sigma^2}=(\sigma_1^2,\cdots,\sigma_q^2)$, ${\bm T}=({\bm T}^{(1)},\cdots,{\bm T}^{(q)})$, and ${\bm\theta}=({\bm\theta}^{(1)},\cdots,{\bm\theta}^{(q)})$.
	There are $1+q+\sum_{j=1}^qm_j(m_j-1)/2+pq$ parameters. To estimate these parameters, Deng et al. (2017) considered the maximum likelihood estimation as
	\begin{align}\label{loglikelihood-1}
		\{\hat{\mu}, \hat{\bm \sigma}^2, \hat{\bm T}, \hat{\bm \theta}\} = \argmax_{\mu, {\bm\sigma^2}, {\bm T}, {\bm\theta}} \left[-\frac 12\log|{\bm\Phi}|-\frac 12({\bm y}_n-\mu{\bm 1}_n)^T{\bm\Phi}^{-1}({\bm y}_n-\mu{\bm 1}_n) \right].
	\end{align}
	Note that matrix ${\bm T}^{(j)}$ needs to be a valid correlation matrix, i.e., ${\bm T}^{(j)}=(\tau_{r,s}^{(j)})_{m_j\times m_j}$ needs to be a positive definite matrix with unit diagonal elements.
	To satisfy this requirement, the hypersphere parameterization approach in Zhou et al. (2011) is adopted here to parameterize ${\bm T}^{(j)}$ for $j =1, \ldots, q$.
	With the estimates obtained from \eqref{loglikelihood-1}, one can calculate $\hat\mu_{0|n}, \hat\sigma^2_{0|n}$ and subsequently compute the predictive distribution of $Y_0|\bm y_{n}$.
	The details of the hypersphere parameterization and the maximum likelihood estimation can be found in the appendix.
	
\section{The Proposed Adaptive CEE Sequential Design}
In this section, we describe the proposed adaptive CEE sequential design based on the additive Gaussian process model for computer experiments with quantitative and qualitative factors.
The proposed adaptive CEE sequential design focuses on the efficient global optimization,
i.e., efficiently finding the optimum through the sequential design procedure.
Without loss of generality, we consider the minimization problem.
That is, given $n$ collected data points $({\bm w}_{t}^T, y_{t}), t=1, \cdots, n$, the key interest is to find the next design point $\bm w_{n+1} \in \mathcal{A}$ for the computer experiment
such that we can promptly find the optimal setting of $\bm w^{*}$ to reach the smallest value of output $y(\bm w)$.
Here $\mathcal{A}$ is the whole design region of $\bm w$, i.e., $\mathcal{A}=\{({\bm x},{\bm z})| {\bm x}\in{\bm X}, {\bm z}\in{\bm Z}\}$.

Specifically, Section 3.1 presents the proposed adaptive CEE method.
Section 3.2 provides some theoretical justification on the adaptiveness regarding the design region for the adaptive CEE method.
In Section 3.3, we make some connections between the proposed adaptive CEE method and the commonly-used EI method for sequential designs.

\subsection{The Adaptive CEE Sequential Design Criterion}
For a computer experiment with quantitative and qualitative factors, suppose that the collected data are $({\bm w}_{t}^T, y_{t}), t=1, \cdots, n$.
We can use the AGP to fit the data and obtain the predictive normal distribution of $Y_0|{\bm y}_n$ for any input $\bm w_{0}$ with mean $\mu_{0|n}({\bm w}_0)$ and variance $\sigma_{0|n}({\bm w}_0)$ as described in \eqref{eq: pred-mean} and \eqref{eq: pred-var}.
To find the design point for minimizing the response output, one would encourage local exploitation, but also has the flexibility of exploration to other regions.
The design point with a small value of $\mu_{0|n}({\bm w}_0)$ will support local exploitation, while the design point with a large value of $\sigma_{0|n}({\bm w}_0)$ will encourage the exploration.
Thus the idea of considering both $\mu_{0|n}({\bm w}_0)$ and $\sigma^{2}_{0|n}({\bm w}_0)$ is natural thinking for choosing the next design point.
Intuitively, we would like to sequentially choose the next point ${\bm w}_0$ when the mean $\mu_{0|n}({\bm w}_0)$ is small and the standard deviation $\sigma_{0|n}({\bm w}_0)$ is large,
which is to achieve a good balance between exploitation and exploration.
Under this consideration, it would be reasonable to consider a criterion of choose the next point ${\bm w}_{n+1}$ as
\begin{align}\label{eq: CEE}
		{\bm w}_{n+1}&=\argmin_{{\bm w}_0 \in \mathcal{A}} \left \{ \hat\mu_{0|n}({\bm w}_0)-\rho \hat\sigma_{0|n}({\bm w}_0) \right \},
\end{align}	
where $\rho \ge 0$ is a tuning parameter. We call such a criterion as the \textit{composite exploitation and exploration} (CEE) criterion.
Note that if $\rho$ is chosen to be $z_{\alpha/2}$, the $\alpha/2$ upper quantile of the standard normal distribution,
then $\hat\mu_{0|n}({\bm w}_0)-\rho\hat\sigma_{0|n}({\bm w}_0)=\hat\mu_{0|n}({\bm w}_0)-z_{\alpha/2}\hat\sigma_{0|n}({\bm w}_0)$ is the lower confidence limit of $[Y_0|{\bm y}_n]$ with the confidence level $1-\alpha$.
It implies that, instead of minimizing the mean of $[Y_0|{\bm y}_n]$,
the CEE criterion is to minimize the lower confidence limit of $[Y_0|{\bm y}_n]$ when searching for the input of achieving the minimum of the response surface.
Therefore, such an expression of the CEE criterion is simple and easy to compute.
Note that the CEE criterion can be easily modified for the maximization problem as $\max_{{\bm w}_0}[\mu_{0|n}({\bm w}_0)+\rho\sigma_{0|n}({\bm w}_0)]$.
	
However, the optimization in \eqref{eq: CEE} requires the search over the whole design space $\mathcal{A}=\{({\bm x},{\bm z})| {\bm x}\in{\bm X}, {\bm z}\in{\bm Z}\}$ in each iteration of the sequential design procedure.
When the inputs of experiments contain both quantitative factors and qualitative factors, the design space $\mathcal{A}$ is discontinuous in nature due to the qualitative factors.
The optimization will be computationally intensive to obtain the global optimum especially when there are a large number of qualitative factors with many levels.
Moreover, as the sequential design procedure is conducted with more collected data points, there should be more information on the design region where the minimum is located.
Therefore, we propose an adaptive CEE criterion for finding the next design point where the design region is adaptive in each iteration of the sequential procedure.
Specifically, the proposed adaptive CEE criterion is to choose the next point ${\bm w}_{n+1}$ as
\begin{align}\label{eq: Adaptive-CEE}
		{\bm w}_{n+1}&=\argmin_{{\bm w_0 \in \mathcal{A}_{n}}} \left \{ \hat\mu_{0|n}({\bm w}_0)-\rho\hat\sigma_{0|n}({\bm w}_0) \right \},
	\end{align}
where $\mathcal{A}_{n} \subset \mathcal{A}$ is the adaptive design region as
\begin{align}\label{eq: Adaptive-Region}
\mathcal{A}_{n} = \left \{{\bm w}_0\in \mathcal{A}:  \hat\mu_{0|n}({\bm w}_0)-\sqrt{\beta_{0|n}} \hat\sigma_{0|n}({\bm w}_0)  \leq  \min_{\bm w_{0}} [ \hat\mu_{0|n}({\bm w}_0)+\sqrt{\beta_{0|n}}\hat\sigma_{0|n}({\bm w}_0)] \right  \}.
	\end{align}
Here is the expression of $\beta_{0|n}$ is specified in Section \ref{sec: justification}.
The justification of choosing $\mathcal{A}_{n}$ in \eqref{eq: Adaptive-Region} will be explained in detail in Theorem~\ref{prop3}. The design region $\mathcal{A}_{n}$ consists of the points $\hat\mu_{0|n}({\bm w}_0)-\sqrt{\beta_{0|n}} \hat\sigma_{0|n}({\bm w}_0)$ is less than the minimum of $\hat\mu_{0|n}({\bm w}_0)+\sqrt{\beta_{0|n}} \hat\sigma_{0|n}({\bm w}_0)$.
\nx{One can see that design region $\mathcal{A}_{n}$, as a subset of the whole region $\mathcal{A}$, varies with the data points collected sequentially. 
In the numerical example in Section 5, we illustrate how $\mathcal{A}_n$ changes as the data arrives. Solving the optimization with the adaptive CEE criterion will be more computationally efficient because the search for next input setting in each iteration is confined in $\mathcal{A}_n$ rather than the whole design region $\mathcal{A}$.}

\subsection{Theoretical Justification for the Adaptive Design Region}\label{sec: justification}
For notation convenience, we use $\mu_{0|n}$ and  $\sigma_{0|n}$ rather than their estimates in the presentation of theoretical investigation and technical proofs.
These theoretical results still hold when the estimates are used. 
The technical proofs can be found in the appendix.  
\nx{Now we focus on finding the adaptive region $\mathcal{A}_{n}$ based on the properties of the predictive mean $\mu_{0|n}({\bm w}_0)$}.
First, we present Lemma~\ref{prop1} below.

\begin{lemma}\label{prop1}
For a given quantitative factors ${\bm x}_0\in{\bm X}$ from a design point ${\bm w}_0 = ( {\bm x}_0^T, {\bm z}_0^T)^T \in \mathcal{A}$, 
let $y({\bm w}_0)$ be a sample from the Gaussian process in \eqref{model}. For all $\alpha \in (0, 1)$, we have
		\begin{align}\label{prop1fo}
			P\left(|y({\bm w}_0)-\mu_{0|n}({\bm w}_0)|\leq \sqrt{\beta_{0|n}}\sigma_{0|n}({\bm w}_0),\forall {\bm z}_0\in {\bm Z}, \forall n \geq 1\right)\geq 1-\alpha,
		\end{align}
where $\beta_{0|n}=2\log(\pi^2n^2M/6\alpha)$ with $M=|{\bm Z}|= \prod_{j=1}^{q} m_{j}$ being the size of ${\bm Z}$, 
and $\mu_{0|n}({\bm w}_0)$ and $ \sigma_{0|n}({\bm w}_0)$ are given in (\ref{eq: pred-mean}) and (\ref{eq: pred-var}).
\end{lemma}

%

Lemma \ref{prop1} is based on Lemma 1 in Jala et al. (2016), which established similar results for a finite space. 
Because $\bm Z$ is a finite discrete space and ${\bm x}_0\in{\bm X}$ is fixed, the design space in Lemma \ref{prop1} is finite.
We  can easily extend their proof to ensure that Lemma \ref{prop1} holds, and thus we skip the proof of Lemma  \ref{prop1}.
	
Lemma \ref{prop1} gives the lower bound and upper bound for the prediction of $y(\bm w_0)$.
Let denote the lower bound $\mu_{0|n}^L({\bm w}_0)$ and the upper bound $\mu_{0|n}^U({\bm w}_0)$ as follows:
	\begin{align}\label{muLU}
		\mu_{0|n}^L({\bm w}_0)=\mu_{0|n}({\bm w}_0)-\sqrt{\beta_{0|n}}\sigma_{0|n}({\bm w}_0),\ \mu_{0|n}^U({\bm w}_0)=\mu_{0|n}({\bm w}_0)+\sqrt{\beta_{0|n}}\sigma_{0|n}({\bm w}_0),
	\end{align}
Then Lemma \ref{prop1} implies that given ${\bm x}_0\in{\bm X}$, $y({\bm w}_0)$ belongs to the interval $[\mu_{0|n}^L({\bm w}_0),\mu_{0|n}^U({\bm w}_0)]$ with the probability greater than $1-\alpha$.
Moreover, Lemma \ref{prop2} below shows that $\min_{\bm w_0\in\mathcal{A}}y({\bm w}_0)$ belongs to the interval  $[\min_{\bm w_0\in\mathcal{A}}\mu_{0|n}^L({\bm w}_0),\min_{\bm w_0\in\mathcal{A}}\mu_{0|n}^U({\bm w}_0)]$ with the probability greater than  $1-\alpha$.
Let us define $y_{\min}, \tilde \mu_{\min,n},  \tilde \mu_{\min,n}^L, \tilde \mu_{\min,n}^U$ as follows:
	\begin{align}\label{eq: esti1-4}
		y_{\min}=\min_{\bm w_0\in\mathcal{A}}y({\bm w}_0), \
		&\tilde\mu_{\min,n}=\min_{\bm w_0\in\mathcal{A}}\mu_{0|n}({\bm w}_0), \nonumber \\
		\tilde\mu_{\min,n}^L=\min_{\bm w_0\in\mathcal{A}}\mu_{0|n}^L({\bm w}_0), \
		&\tilde\mu_{\min,n}^U=\min_{\bm w_0\in\mathcal{A}}\mu_{0|n}^U({\bm w}_0).
	\end{align}
%
\begin{lemma}\label{prop2}
		Let $y_{\min}, \tilde\mu_{\min,n}^L, \tilde\mu_{\min,n}^U$  be the quantifies as defined in \eqref{eq: esti1-4}. Then for all $\alpha\in(0,1)$,
		\begin{align}
			P\left(y_{\min}\in[\tilde\mu_{\min,n}^L, \tilde \mu_{\min,n}^U],\forall n\geq 1\right)\geq 1-\alpha.
		\end{align}
\end{lemma}
The proof of Lemma \ref{prop2} can be found in the appendix.
Now we can obtain the bound for the discrepancy between of the minimum of the response $y_{\min}$ and its estimate $\tilde\mu_{\min,n}$ in a probabilistic manner.
\begin{theorem}\label{prop3}
		Let $\mu_{0|n}^L({\bm w}_0), y_{\min}, \tilde\mu_{\min,n}, \tilde\mu_{\min,n}^U$ be the quantitie as defined in \eqref{muLU} and \eqref{eq: esti1-4}.
Then for all $\alpha\in(0,1)$, we have
        \begin{align}
			P\left(|\tilde\mu_{\min,n}-y_{\min}|\leq \sqrt{\beta_{0|n}}\sup_{{\bm w}_0\in \mathcal{A}_{n}}\sigma_{0|n}({\bm w}_0),\forall n\geq 1\right)\geq 1-\alpha,
		\end{align}
where $\mathcal{A}_{n} = \{{\bm w}_0\in \mathcal{A}: \mu_{0|n}^L({\bm w}_0)\leq \tilde\mu_{\min,n}^U\}$.
	\end{theorem}
Clearly, the definition of $\mathcal{A}_{n}$ here is the same as in  \eqref{eq: Adaptive-CEE}.
It is easy to see that the lower bound $\mu^L_{0|n}(\bm w_0)$ is smaller than the upper bound $\mu^U_{0|n}(\bm w_0)$ for every $\bm w_0\in\mathcal{A}$.
In Theorem \ref{prop3}, the design region $\mathcal{A}_{n}$ consists of the points whose the lower bound $\mu^L_{0|n}(\bm w_0)$ is less than the minimum of its upper bound $\tilde \mu^U_{\min,n}$.
Thus $\mathcal{A}_{n}$  can have a smaller size than the whole design region $\mathcal{A}$.
Theorem \ref{prop3} provides a bound for the difference between $y_{\min}$ and its estimate $\tilde\mu_{\min,n}$, which depends on $\mathcal{A}_{n}$. 
It implies that  $\tilde\mu_{\min,n}$ will be in a small neighborhood of $y_{\min}$ with a relatively high probability.
\nx{Furthermore, Corollary \ref{prop4} below shows the adaptive design region $\mathcal{A}_{n}$ will cover the true optimal setting ${\bm w}^* = \arg \min_{\bm w} y({\bm w})$  with a high probability.}
\begin{corollary}
        \label{prop4}	
		Denote ${\bm w}^*$ be one of the optimal points that minimize $y({\bm w})$,
i.e., $y({\bm w}^*) = \min_{{\bm w}\in \mathcal{A}} y({\bm w})$.
Then for all $\alpha\in(0,1)$, we have
		\begin{align*}
			P\left(\mathcal{A}_{n} \ni {\bm w}^*, \forall n \geq 1\right)\geq 1-\alpha,
		\end{align*}
where $\mathcal{A}_{n}$ is the design region defined in Theorem \ref{prop3}.
\end{corollary}

\subsection{Connection to the Expected Improvement Criterion}
This section provides some connections between the proposed adaptive CEE criterion and the EI criterion for the sequential design of computer experiments with quantitative and qualitative factors. 
In fact, the EI criterion was usually used for computer experiment with only quantitative inputs, but it can work for computer experiment with mixed inputs.
	
Let $I({\bm w})$ be an improvement function defined for any ${\bm w}=({\bm x}^T, {\bm z}^T)^T$ with quantitative factors ${\bm x}$ and qualitative factors ${\bm z}$ in the input space. The corresponding EI criterion is given by the expectation of $I({\bm w})$:
	\begin{align*}
		E[I({\bm w})]=\int_{y\in R} I({\bm w})f(y|\bm w, {\bm y}_n)dy,
	\end{align*}
	where $f(y|\bm w, {\bm y}_n)$ is the predictive density of $Y$ conditional on all runs so far.
	
	For the minimization problem, Jones et al. (1998) proposed the improvement function
	\begin{align*}
		I({\bm w}_0)=\max\left\{y_{\min,n}-Y({\bm w}_0),0\right\},
	\end{align*}
	where $y_{\min,n}$ is the minimum value of the obtained responses among the $n$ runs.
By \eqref{eq: pred-mean} and \eqref{eq: pred-var}, the closed form of the EI can be expressed as
\begin{align}\label{EIc}
\small E[I({\bm w}_0)]=\sigma_{0|n}({\bm w}_0)\phi\left(\frac{y_{\min,n}-\mu_{0|n}({\bm w}_0)}{\sigma_{0|n}({\bm w}_0)}\right)+\left(y_{\min,n}-\mu_{0|n}({\bm w}_0)\right)\Phi\left(\frac{y_{\min,n}-\mu_{0|n}({\bm w}_0)}{\sigma_{0|n}({\bm w}_0)}\right),
	\end{align}
	where $\phi$ and $\Phi$ denote the probability density and the cumulative distribution of the standard normal distribution.
The EI method is to choose  ${\bm w}_{n+1}$ to be the maximizer of $E[I({\bm w}_0)]$,
	\begin{align}\label{EI}
		{\bm w}_{n+1}=\argmax_{{\bm w}_0}\ E[I({\bm w}_0)].
	\end{align}
Jones et al. (1998) stated the derivative of the expected improvement in \eqref{EIc} with respect to $\mu_{0|n}({\bm w}_0)$ or $\sigma_{0|n}({\bm w}_0)$ results in the surprisingly simple expressions:
	\begin{align*}
		&\frac{\partial E[I({\bm w}_0)]}{\partial\mu_{0|n}({\bm w}_0)}=-\Phi\left(\frac{y^{(n)}_{\min}-\mu_{0|n}({\bm w}_0)}{\sigma_{0|n}({\bm w}_0)}\right) \le 0,\\
		&\frac{\partial E[I({\bm w}_0)]}{\partial\sigma_{0|n}({\bm w}_0)}=\phi\left(\frac{y^{(n)}_{\min}-\mu_{0|n}({\bm w}_0)}{\sigma_{0|n}({\bm w}_0)}\right) \ge 0.
	\end{align*}
Thus, the objective function $E[I({\bm w}_0)]$  in \eqref{EIc} is larger when $\mu_{0|n}({\bm w}_0)$ is lower and $\sigma_{0|n}({\bm w}_0)$ is higher.
Such conditions are consistent with the motivation of the proposed CEE method.
Note that the EI criterion involves the normal probability density function $\phi$ and the cumulative density function $\Phi$, while the proposed CEE has a much simpler expression with meaningful interpretation. The adaptive design region of the proposed method also makes the search of the next design point more efficient. 


	
	\section{Numerical Examples}
	In this section, we investigate the performance of the proposed adaptive CEE method in comparison with the four benchmark methods defined as 
	\begin{itemize}
		\item[(i)] EI: the method sequentially maximizes the expected improvement in \eqref{EI} as acquisition function;
		
		\item[(ii)] MU: the method sequentially minimizes the prediction mean as acquisition function, i.e., $ {\bm w}_{n+1}=\argmin_{{\bm w}_0}\ \hat\mu_{0|n}({\bm w}_0)$;
		
		\item[(iii)] SI: the method sequentially maximizes the prediction variance as acquisition function, i.e., ${\bm w}_{n+1}=\argmax_{{\bm w}_0}\ \hat\sigma_{0|n}({\bm w}_0)$;
		
		\item[(iv)]  RA: the one-shot design approach.
	\end{itemize}
The first three approaches are sequential designs, each of which chooses the next design point by the given method, gets its response and updates the model estimation, and then continues to choose the next design point until the stopping criterion is met. Here we consider the methods in comparison have the same number of runs. 
In each numerical example, we will report the minimal values found by the five methods in comparison.
%
	
\subsection{Example with a Qualitative Factor and a Quantitative Factor}
	\begin{example}
		Consider the simple case that there is only one quantitative factor $x\in[0,1]$ and one qualitative factor $z$ of three levels.
		The underlying function for the output response $y$ is expressed as
		\begin{align}\label{Ex1}
			y=\left\{
			\begin{array}{ll}
				\vspace{0.2cm}
				2+\cos(6\pi x),&{\rm if}\  z=1,\\
				\vspace{0.2cm}
				1-\cos(4\pi x),& {\rm if}\ z=2,\\
				\cos(2\pi x),& {\rm if}\ z=3.
			\end{array}\right.
		\end{align}
It is easy to see that the minimum of the function in \eqref{Ex1} is obtained exactly at $z=3$ and $x=0.5$.
	\end{example}
	
To start the proposed adaptive CEE sequential design, we obtain an initial training data of three points,  
where a three-level full factorial design (Wu and Hamada 2009) is used for the qualitative factor and a random Latin hypercube design (McKay et al. 1979) is used for the quantitative factor.
In each iteration of the sequential design, the corresponding output value of the chosen design point is calculated by \eqref{Ex1},
and the minimum of the obtained output values is regarded as the minimum of \eqref{Ex1}.
For the proposed adaptive CEE method, 
we choose $\rho=2$. 
We have compared the proposed method with different values of $\rho=0.5,1,2,3$. 
The results appear to have similar performance.
When we choose $\rho=2$, $\hat\mu_{0|n}({\bm w}_0)-\rho\hat\sigma_{0|n}({\bm w}_0)=\hat\mu_{0|n}({\bm w}_0)-2\hat\sigma_{0|n}({\bm w}_0)$ can be viewed as the lower confidence limit of $Y_0|\bm y_n$ with confidence level around $0.95$.  
In order to obtain the minimum of the response, the proposed method is to minimize the lower confidence limit of $Y_0|\bm y_n$. 
Thus it is more reasonable than minimizing the mean of $Y_0|\bm y_n$. 
Hereafter, we choose $\rho=2$ in the simulation.

	
	
	\begin{figure}[ht]
		\centering
		\resizebox{16cm}{8cm} {\includegraphics[width=90mm]{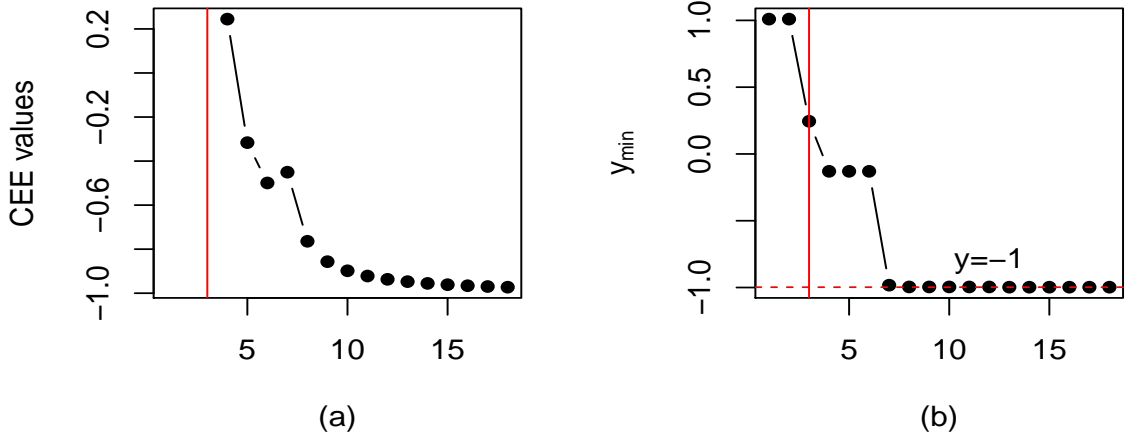}}
		\caption{Results of the adaptive CEE sequential design in one simulation trial, (a) the CEE value, (b) the obtained minimum of response, where 15 points are selected sequentially based on three initial runs.}\label{figure1-line}
	\end{figure}

Figure \ref{figure1-line} shows the results of the adaptive CEE sequential design in one simulation, 
where the three initial points and $15$ sequentially added points are on the left and right of the red vertical line respectively.
Here the CEE value in Figure \ref{figure1-line}(a) represents the value of $\hat\mu_{0|n}({\bm w}_{n+1})-\rho\hat\sigma_{0|n}({\bm w}_{n+1})$ in each iteration.
It is seen that the CEE value converges quickly within 10 iterations of the sequential runs.
From Figure \ref{figure1-line}(b), 
it is clear that the estimated minimum of the responses drops sharply as the points are added sequentially by the proposed adaptive CEE method.
	
\begin{figure}
		\centering
		\resizebox{10cm}{8cm} {\includegraphics[width=90mm]{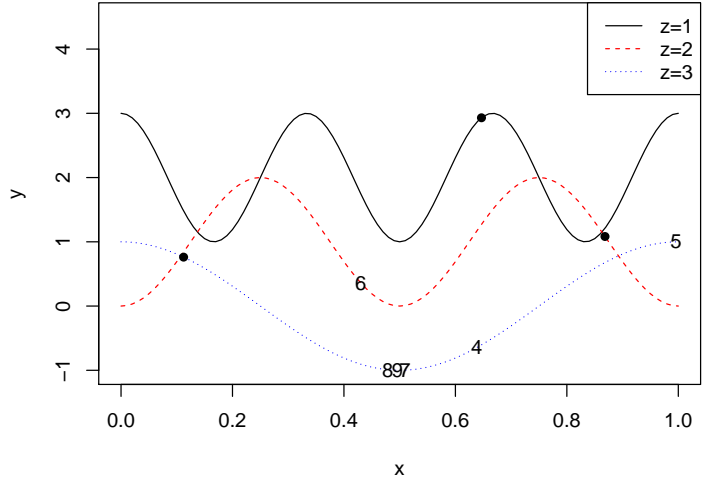}}
		\caption{Illustration of the adaptive CEE sequential inputs with three initial points and six sequentially added points.}\label{figure1-point}
\end{figure}

Note that the true minimum of the function in \eqref{Ex1} is $-1$. 
The proposed adaptive CEE method achieves the minimum with four iterations of the sequential procedure. Moreover, when the minimum is achieved, the sequential inputs converge at the minimum point.
Figure \ref{figure1-point} marks the selected design points corresponding to the simulation in Figure \ref{figure1-line}.
In Figure \ref{figure1-point}, the small solid dots are the initial three points, and the points which are labeled ``4" to ``9" are six sequential points.
From Figure \ref{figure1-point}, the proposed method efficiently allocates the design point to the level $z=3$ to seek the minimum of response.
The points marked with ``7", ``8", ``9" almost coincide at $z=3$, with their responses values close to the true minimal response value of $-1$.
	
\begin{figure}[ht]
		\centering
		\resizebox{15cm}{12cm} {\includegraphics[width=90mm]{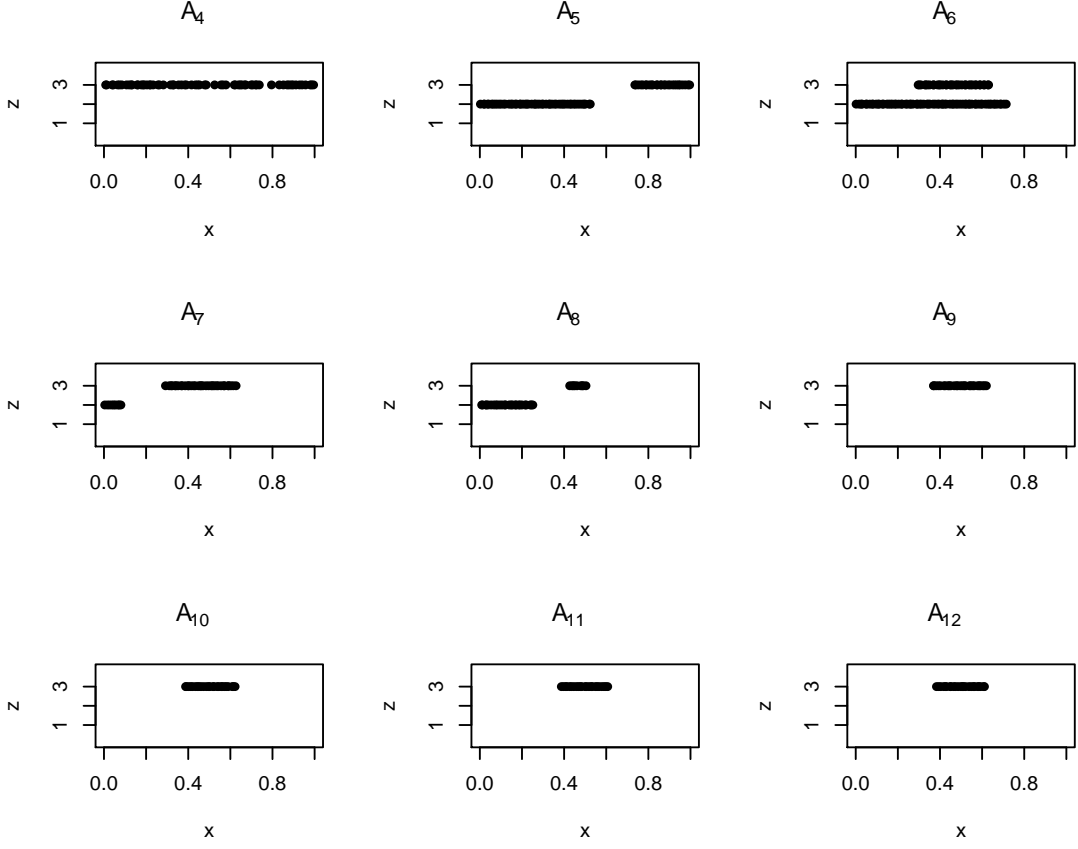}}
		\caption{Illustration of the adaptive $\mathcal{A}_{n}$ design region from one simulation in Example 1.}\label{figure1-L}
\end{figure}
	
To further examine the performance of the proposed adaptive CEE method, 
it is of interest to understand how the adaptive set of the feasible region behaves.
Figure \ref{figure1-L} reports the sequential $\mathcal{A}_{n}$ subsets corresponding to the sequential design in Figures \ref{figure1-point} and \ref{figure1-line}.
From Figure \ref{figure1-L}, one can clearly observe that $\mathcal{A}_{n}$ quickly converges to the set $\{(x,z):x\in(0.44,0.66),z=3\}$.
	Note that the true minimum point $(0.5,3)$ belongs to this set. 
	It provides further evidence for the theoretical investigation in Theorem \ref{prop4} regarding to the convergence properties of the proposed method.
	
\begin{figure}[ht]
		\begin{center}
			\resizebox{12cm}{8cm}{\includegraphics[width=80mm]{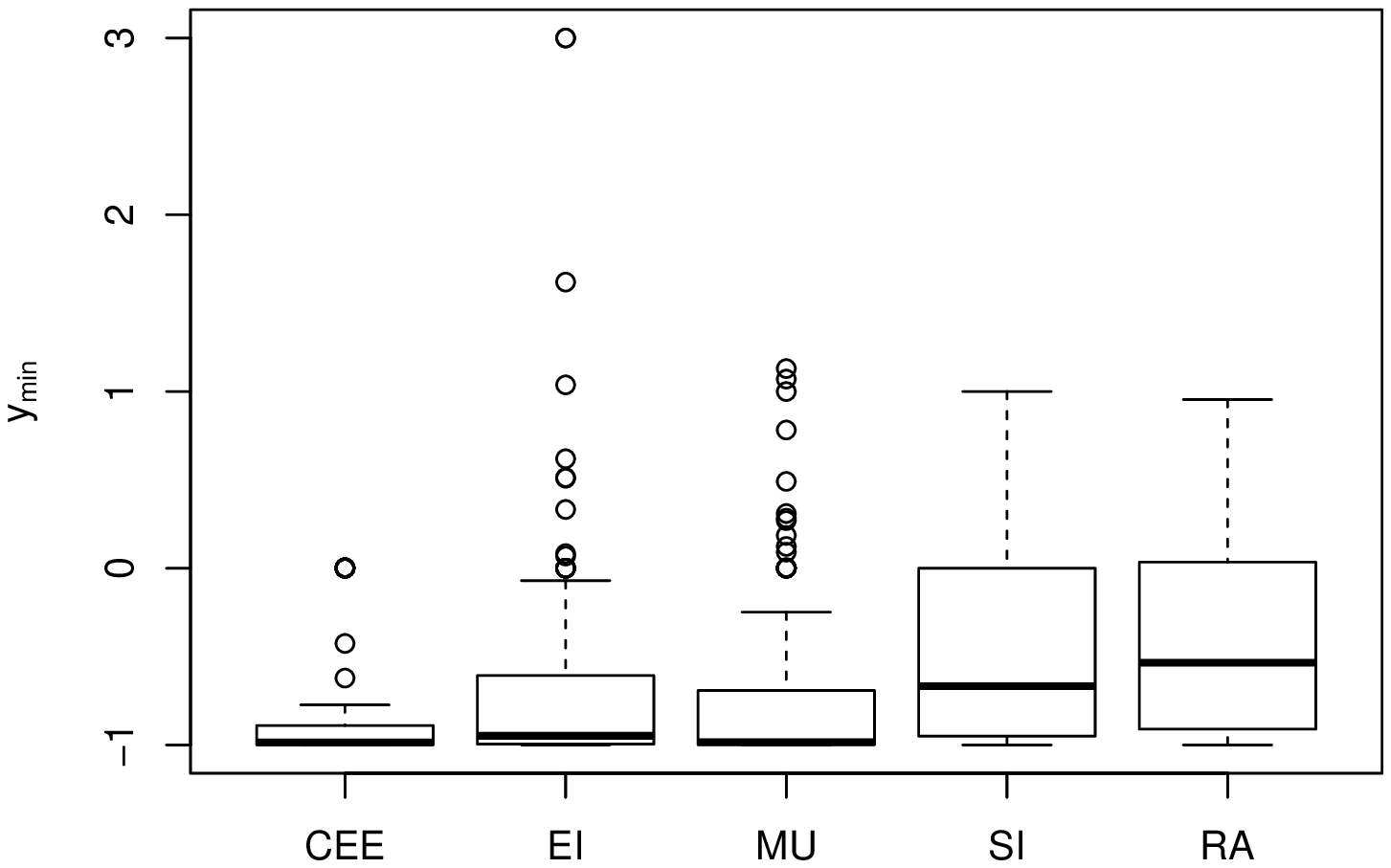}}\\ 
			\resizebox{16cm}{4cm}{\includegraphics[width=90mm]{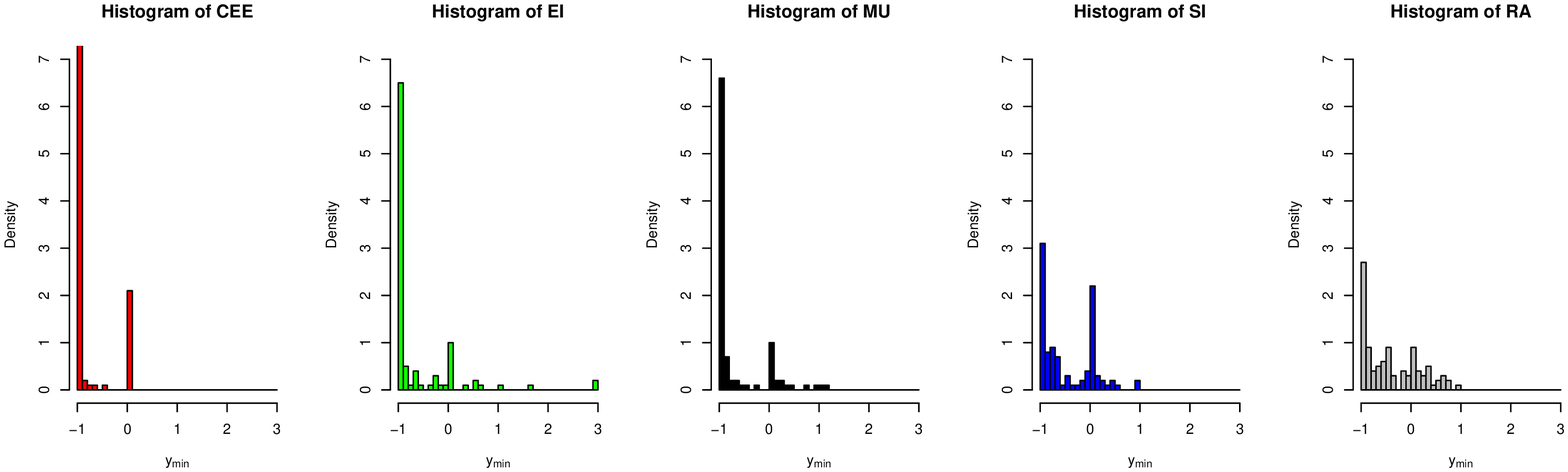}}
			\caption{Boxplots and histograms of the obtained minimum values of response over 100 simulations in Example 1.}
			\label{figure1-box}
		\end{center}
\end{figure}

Now we compare the proposed adaptive CEE method with the other four benchmark methods over 100 simulations.
For each method, the same number of initial run of size three is used with a three-level full factorial design for the qualitative factor and a random Latin hypercube design is for the quantitative factor. Six sequential points are chosen for each method in comparison.
Figure \ref{figure1-box} reports the boxplots and histograms of the obtained minimums for methods in comparison.
From Figure \ref{figure1-box}, it is seen that the performance of the adaptive CEE sequential design is much better than the EI, MU, SI and RA designs.
For the proposed adaptive CEE method, most minimums are near $-1$.
For the same number of sequential runs, the adaptive CEE method can obtain the true minimum with a higher probability than the EI, MU, SI, and RA methods.
	
	
\subsection{Examples with Multiple Quantitative and Qualitative Factors}
	\begin{example}
This example is used in Deng et al. (2017) for a computer experiment with $p=3$ quantitative factors and $q=3$ qualitative factors.
The response of the experiment has the following expression
		\begin{align}\label{Ex33}
			y=\sum_{i=1}^3\frac{x_iz_{4-i}}{4000}+\prod_{i=1}^3\cos\left(\frac{x_i}{\sqrt{i}}\right)\sin\left(\frac{z_{4-i}}{\sqrt{i}}\right),
		\end{align}
		where $-100<x_i<100$ for $i=1,\cdots,p$ and $z_j=\{-50,0,50\}$ for $j=1,\cdots,q$.
	\end{example}

\begin{figure}
		\begin{center}
			\resizebox{12cm}{8cm}{\includegraphics[width=80mm]{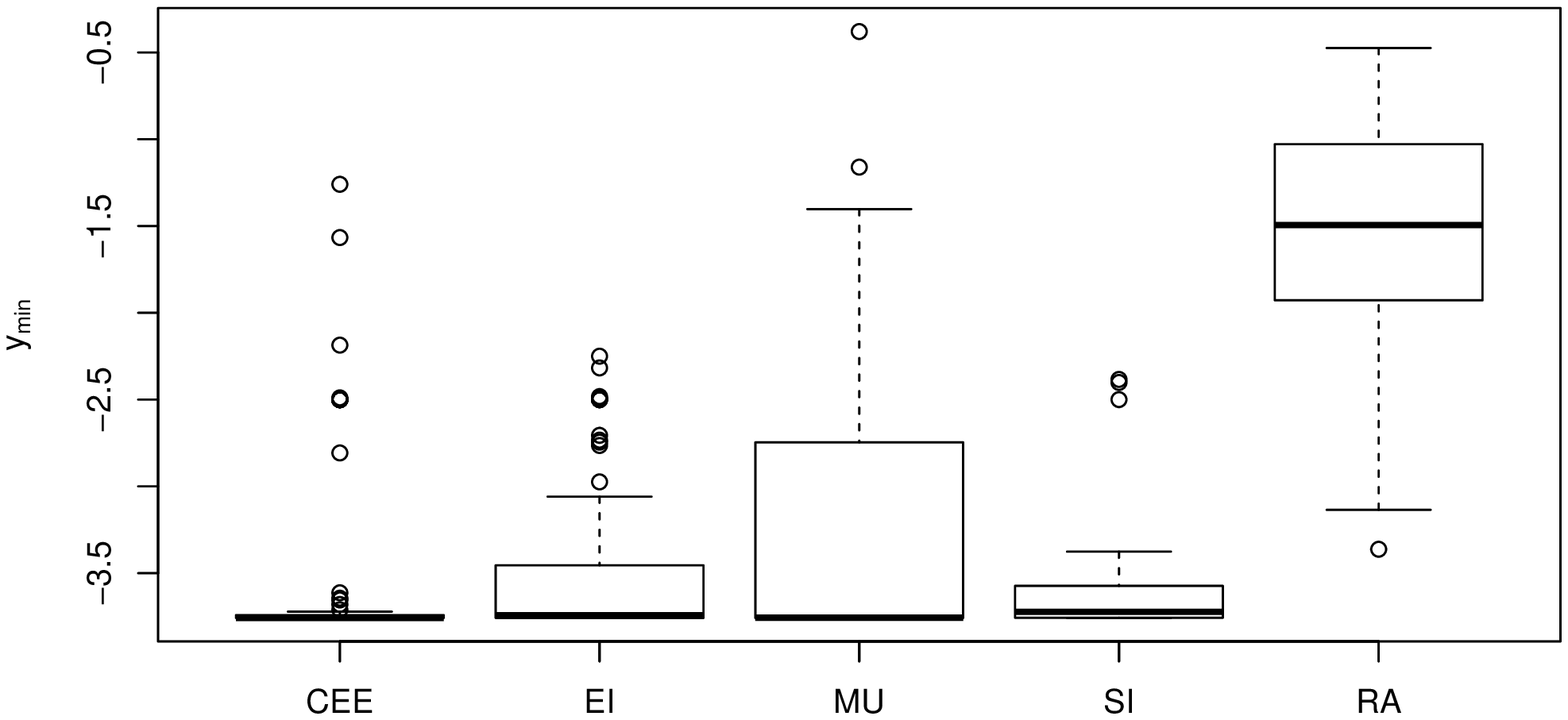}}\\ 
			\resizebox{16cm}{4cm}{\includegraphics[width=90mm]{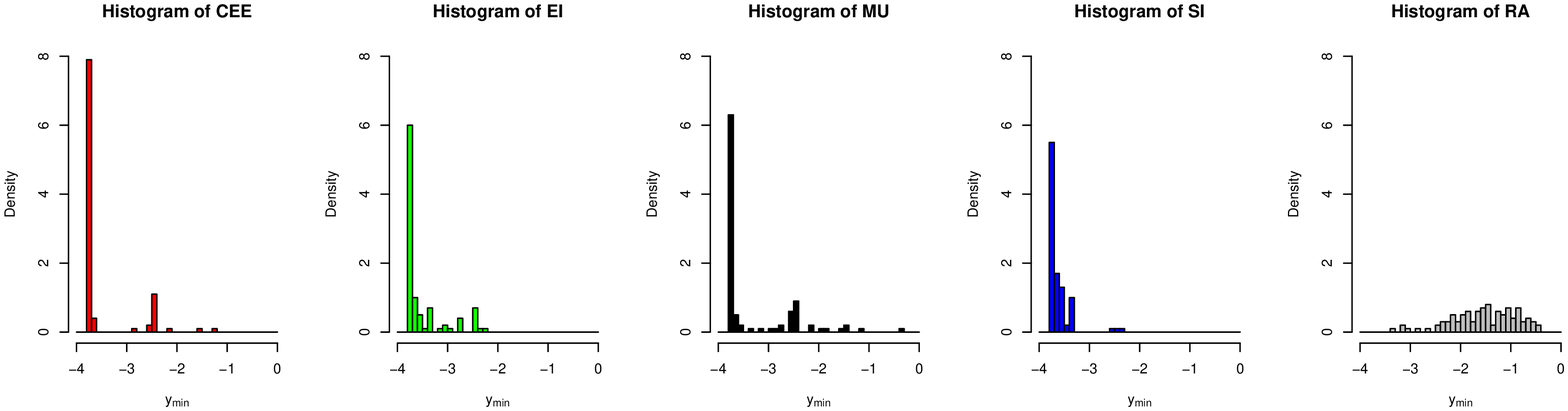}}
			\caption{Boxplots and histograms of the obtained minimums of the response over 100 simulations in Example 2.}
			\label{figure3-box}
		\end{center}
	\end{figure}

Note that the qualitative factors in this example behave as the ordinal factors. 
In each simulation, a 9-run initial design is adopted, 
where a three-level fractional factorial design is used for the qualitative factors and a random Latin hypercube design is used for the quantitative factors.
It is easy to know that the true minimum of \eqref{Ex33} is $3.75$.
The EI, MU, SI and RA designs are compared with the proposed adaptive CEE design. 
For the proposed adaptive CEE design, we choose $\rho=2$.
For each method in comparison, it has the same number of initial runs and then nine follow-up points are obtained sequentially.
Figure \ref{figure3-box} displays the boxplots and histograms of minimums over 100 simulations.
From Figure \ref{figure3-box}, it is seen that the proposed adaptive CEE method outperforms the EI, MU, SI and RA methods significantly in terms of the obtained minimal values.
Over 100 simulations, most minimums found by the adaptive CEE method are very close to the true minimum $3.75$.

\begin{example}
Consider a computer experiment with $p = 3$ quantitative factors and $q = 3$ qualitative factors. 
The response is
$$
y = f_i({\bf x}) \times ( g_j({\bf x}) + h_k({\bf x})),
$$
where $i$, $j$, $k$ are the levels for the first, second and third qualitative factors, $0 \leqslant x_i \leqslant 1$ for $i = 1,2,3$. 
The functions $f_i$, $g_j$ and $h_k$ have the expression as follows:
\begin{align*}
 f_1(\bf x) &= x_1 + x_2^2 +x_3^3, \ f_2({\bf x}) = x_1^2 + x_2 +x_3^3, \\
 f_3(\bf x) &= x_1^3 + x_2^2 +x_3, \  g_1({\bf x}) = \cos(x_1) + \cos(2x_2) +\cos(3x_3),\\
 g_2(\bf x) &= \cos(3x_1) + \cos(2x_2) +\cos(x_3), \ g_3({\bf x}) = \cos(2x_1) + \cos(x_2) +\cos(3x_3),\\
 h_1(\bf x) &= \sin(x_1) + \sin(2x_2) + \sin(3x_3), \  h_2({\bf x}) = \sin(3x_1) + \sin(2x_2) + \sin(x_3),\\
 h_3(\bf x) &= \sin(2x_1) + \sin(x_2) + \sin(3x_3).
\end{align*}
\end{example}

\begin{figure}[hb]
\begin{center}
\begin{tabular}{cc}
\includegraphics[width=0.4\textwidth]{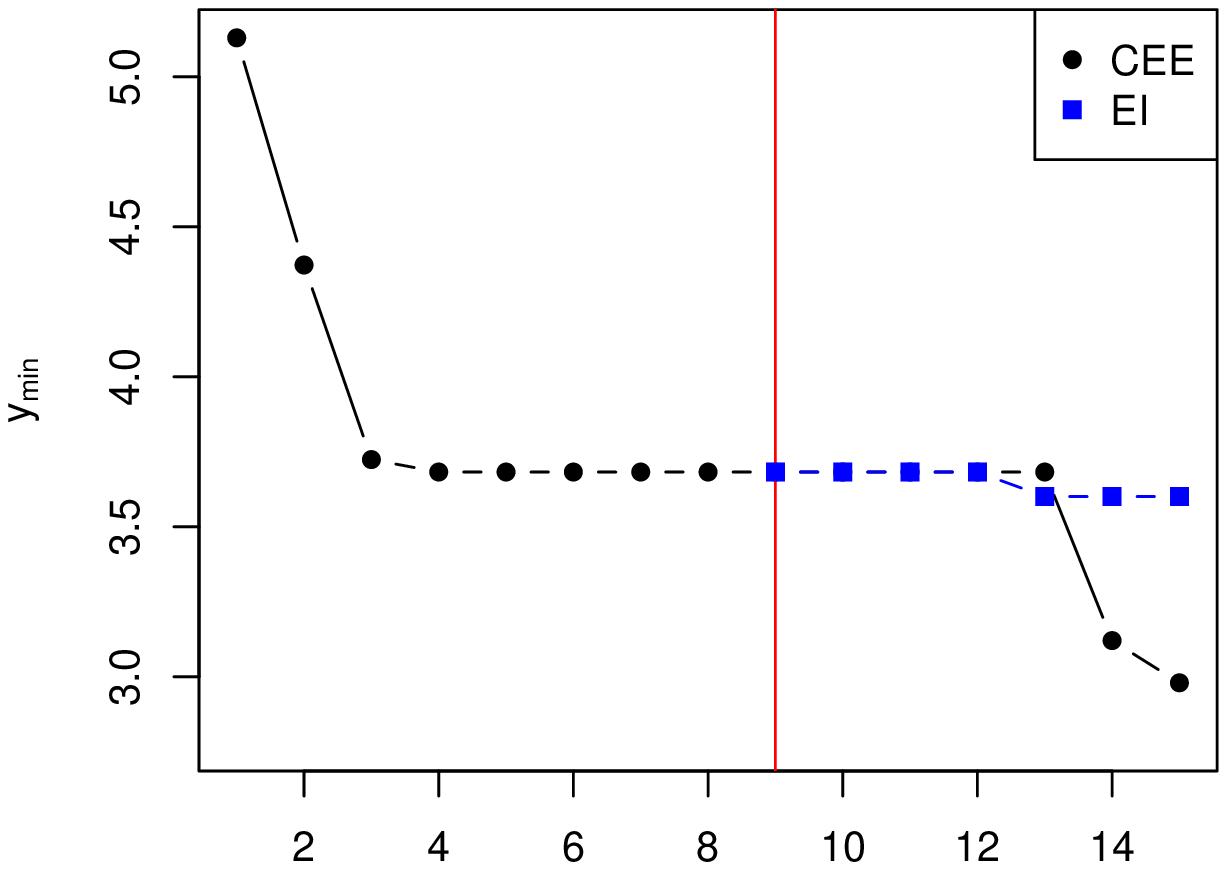}&
\includegraphics[width=0.4\textwidth]{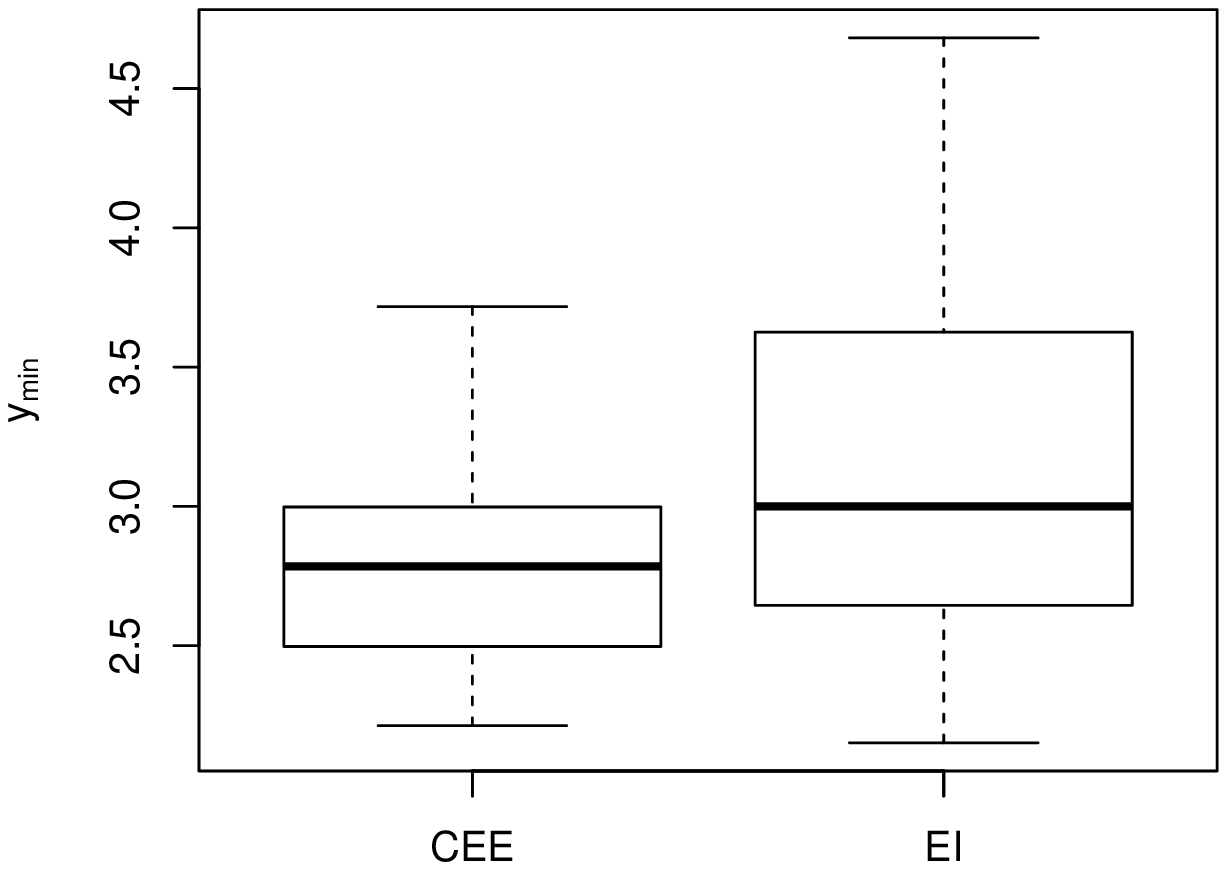}\\[-2ex]
(a) & (b)
\end{tabular}
\caption{The performance of the proposed adaptive CEE method and EI methods for Example 3: (a) the sequential procedure in one simulation; (b) the boxplot of the obtained minimums of the response over 100 simulations.}\label{figure3}
\end{center}
\end{figure}	
	
For this example, a 9-run initial design is adopted, where a three-level fractional factorial design is used for the qualitative factors and a random Latin hypercube design is used for quantitative factors. Six follow-up sequential points are selected by each design in comparison, respectively.
Here, we focus on the comparison between the proposed adaptive CEE method and the EI method as the EI method is commonly used in the literature with appealing performance.
For the proposed adaptive CEE design, we choose $\rho=2$.
Figure \ref{figure3}(a) reports the minimums of the adaptive CEE sequential design in comparison with the EI design in one simulation.
Figure \ref{figure3}(b) shows the boxplots of the values of minimums obtained by the adaptive CEE sequential design in comparison with the EI design in 100 simulations.
It is seen that the proposed adaptive CEE method is much more efficient than the EI method in finding the minimal response output.

\section{Case Study of HPC Data}

In this section, we apply the proposed adaptive CEE sequential design for studying the HPC systems, 
which are important infrastructures to advance the industry 4.0. 
To enhance the performance of HPC system, a key step is to understand the HPC variability since there are run-to-run variation in the execution of a computing task.
In particular, the input/output (IO) throughput (i.e., data transfer speed) is an important metric, which is affected by various system factors such as CPU frequency, the number of threads, IO operation mode, and IO scheduler.
The relationship between the IO throughput ($y$) and these system factors can be quite complicated. 
Moreover, some of these system factors are quantitative and some are qualitative.

In this case study, our objective is to find an optimal level combination of system factors that optimizes the IO performance variability measure.
Table~\ref{tab:exp.setup} summarizes the input factors, of which the quantitative factors are the CPU clock frequency ($x_1$) and the number of threads ($x_2$), and the qualitative factors are the IO operation mode ($z_1$) with three levels, the IO scheduler ($z_2$) with three levels and the VM IO scheduler ($z_3$) with three levels.
Here the IO scheduler is the method that computer operating systems use to decide in which order the block IO operations will be submitted to storage volumes. 
For the IO operation modes, Initialwrite measures the performance of writing a new file, 
Randomread measures the performance of reading a file with accesses being made to random locations within the file, 
and Fwrite measures the performance of writing a file using fwrite() function.
The HPC server is configured with a dedicated 2TB HDD on a 2 socket, 4 core (2 hyperthreads/core) Intel Xeon E5-2623 v3 (Haswell) platform with 32 GB DDR4, using Linux operating system.
The IOzone benchmark task (Norcott 2020) was used in this computer experiment (Xu et al. 2020). 

For a given level combination of input factors as a configuration,
the HPC server executes the IOzone benchmark task and the IO throughput (in kilobytes per second) is recorded.
By executing for 40 replicates, the mean and the standard deviation (SD) of the 40-replicate IO throughput values are calculated (Cameral et al. 2009).
Clearly, a smaller value of the SD indicates the robustness of the HPC system, and a large mean value indicates the effectiveness of the HPC system.
Hence, we consider to use the signal-to-noise ratio $Y_{SN}$, i.e., the ratio of the mean and SD of the throughput values, as the output response in the optimization.

	\begin{table}
		\begin{center}
			\caption{A summary of input factors in the IO throughput experiment of the HPC system.}\label{tab:exp.setup}
			\vspace{.5em}
			\begin{tabular}{rlcc}
			\hline \hline
				\multirow{2}{*}{Category}      & \multirow{2}{*}{Variable} & No. of  & \multirow{2}{*}{Values}\\
				&                           & levels  &                        \\\hline
				Hardware      & $x_1$: CPU clock frequency      & \multirow{2}{*}{continuous} & 1.2, 1.4, 1.5, 1.6, 1.8, 1.9, 2.0, 2.1 \\
				&    (GHz)                        &    & 2.3, 2.4, 2.5, 2.7, 2.8, 2.9, 3.0\\\hline
				Operating     & $z_2$: IO scheduler         & 3  & CFQ, DEAD, NOOP \\
				System        & $z_3$: VM IO scheduler        & 3  & CFQ, DEAD, NOOP\\\hline
				\multirow{2}{*}{Application}& $z_1$: IO operation mode   & 3 & Fwrite, Initialwrite, Randomread\\
				& $x_2$: number of threads  & continuous  & 1,   2,   4,   8,  16,  32,  64, 128, 256\\
				\hline \hline
			\end{tabular}
		\end{center}
	\end{table}

\begin{figure}[ht]
\begin{center}
\begin{tabular}{cc}
\includegraphics[width=0.5\textwidth]{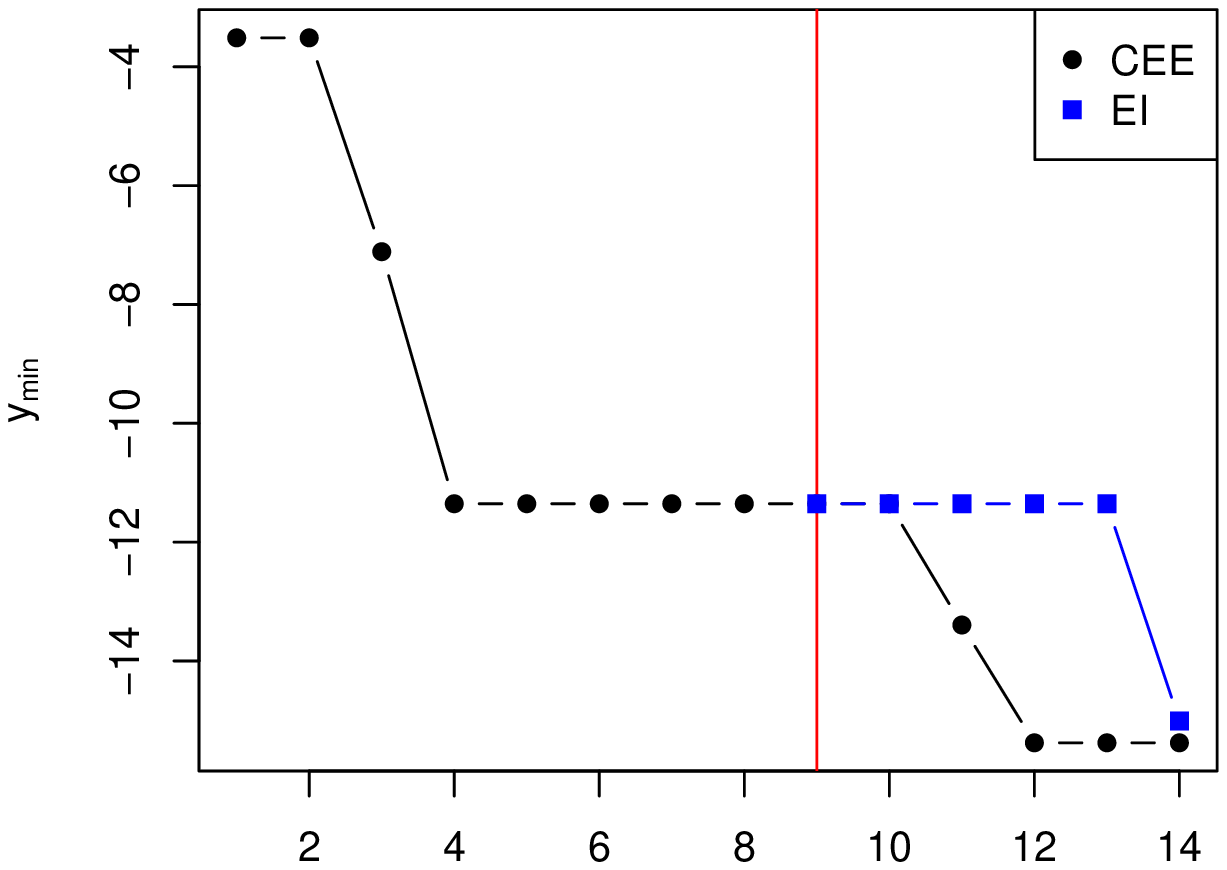}&
\includegraphics[width=0.44\textwidth]{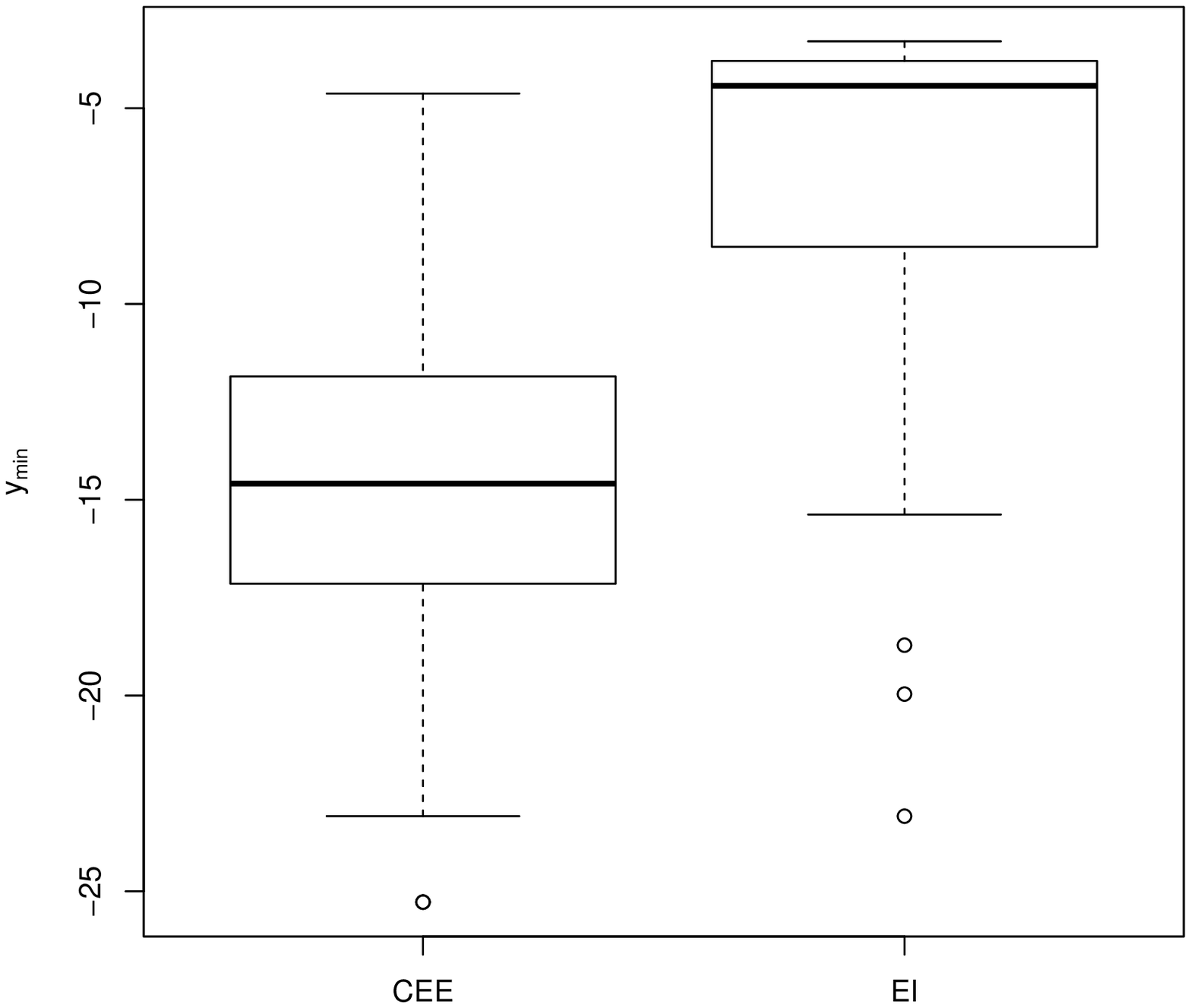}\\[-2ex]
(a) & (b)
\end{tabular}
\caption{The performance of the proposed adaptive CEE method and EI methods for the HPC case study: 
(a) the sequential procedure in one simulation; (b) the boxplot of the obtained minimums of the response ($-Y_{SN}$) over 100 simulations.}\label{Iozone1}
\end{center}
\end{figure}

We apply the proposed adaptive CEE sequential design to find the optimal configuration to achieve the maximum of $Y_{SN}$, the ratio between the mean and SD of the throughput values.
It appears that there is little domain knowledge on the configuration (the setting of input factors) to maximize the ratio of the mean and standard deviation of the throughput values.
Note that maximizing $Y_{SN}$ is equivalent to minimizing $-Y_{SN}$. Thus we use $-Y_{SN}$ as the response for our proposed adaptive CEE sequential design.
For the initial experiment, we consider a 9-run design with a three-level fractional factorial design for the qualitative factors and a random Latin hypercube design for quantitative factors.
Then five design points are obtained sequentially by each method in comparison.
Here, we focus on the comparison between the proposed adaptive CEE sequential and the EI method.
For the proposed adaptive CEE design, we choose $\rho=2$.
Figure \ref{Iozone1}(a) reports the obtained minimums of $-Y_{SN}$ in one simulation trial with the same initial runs from the adaptive CEE method and the EI method, respectively.
One can see that the adaptive CEE sequential design performs much better than the EI design. 
The adaptive CEE method obtains a smaller value of response than the EI method under the same number of runs. 
In the case of Figure \ref{Iozone1}(a), the maximum of $Y_{SN}$ obtained by the proposed adaptive CEE method is $20.198$ at the setting $x_1=1.2$, $x_2=2$, $z_1=``\textrm{Initialwrite}"$, $z_2=``\textrm{NOOP}"$, $z_3=``\textrm{NOOP}"$.
It is interesting to note that the number of threads in this optimal setting of maximizing $Y_{SN}$ is $x_{2} = 2$.
For different initial runs, it happens that the adaptive CEE method performs better than the EI method in most cases.
Moreover, we also compare the adaptive CEE sequential design with the EI design in 100 simulations with different initial runs.
Figure \ref{Iozone1}(b) reports the boxplots of minimal response ($-Y_{SN}$) values obtained by the adaptive CEE method in comparison with the EI method.
Clearly, the proposed adaptive CEE sequential design is much more efficient than the EI design in finding the maximal value of $Y_{SN}$.

\section{Discussion}
In this work, we propose an adaptive CEE sequential design for optimization of computer experiments with qualitative and quantitative factors.
Here we have focused on finding the optimal level combination of factors to minimize (or maximize) the response output.
The proposed adaptive CEE sequential design combines the predictive mean and standard derivation to achieve the balance between exploitation and exploration with meaninful interpretation. Moreover, the adaptive design region varies with the collected design points with a theoretical justification based on the bound between the true optimal response and its estimate.
The proposed method is applied to HPC performance optimization to successfully find the good setting for enhancing the IO throughput.  

Currently the proposed adaptive CEE sequential design criterion is built based on the additive Gaussian process (AGP) in Deng et al. (2017) for computer experiment with qualitative and quantitative factors. The proposed methodology can also be applicable for other Gaussian process models for computer experiment with qualitative and quantitative factors.
It is worth pointing out that the theoretical justification of the adaptive design region can be extend to the case of contour estimation for computer experiment with qualitative and quantitative factors. 

There are several directions for the further development of the proposed method. First, when the number of input variables or the levels of qualitative factors is large, 
The use of AGP can involve a large number of parameters and consequently the sequential procedure can be slow due to the parameter estimation and the large number of possible design points. One possible solution to overcome this drawback is to adopt a more parsimonious Gaussian process model (Zhang et al. 2020). 
To address the intensive search over a large design region in the optimization in \eqref{eq: Adaptive-CEE}, more advanced optimization technique is needed such as the mixed integer programming used in Xie and Deng (2020). Second, there is a tuning parameter $\rho$ in the proposed adaptive CEE criterion in \eqref{eq: Adaptive-CEE}. 
In Section 4.1, the numerical results shown that there is no significant difference on the performance of the proposed method under $\rho=0.5,1,2,3$. 
It will be interesting to understand the relationship between an appropriate choice of the tuning parameter and the choice of the adaptive design region.
Finally, the proposed method is not limited to the continuous response. It will be interesting to investigate on how to extend the proposed method for computer experiments with non-continuous output such as binary responses (Sung et al. 2020).    

\section*{Appendix}
	{\bf Details of the Maximum Likelihood Estimation of AGP}\\
	Under the AGP model (\ref{model}), the log-likelihood function is
	\begin{align}\label{loglikelihood}
		l(\mu, {\bm\sigma^2}, {\bm T}, {\bm\theta})=-\frac n2\log(2\pi)-\frac 12\log|{\bm\Phi}|-\frac 12({\bm y}_n-\mu{\bm 1}_n)^T{\bm\Phi}^{-1}({\bm y}_n-\mu{\bm 1}_n).
	\end{align}
	Setting the derivative of $l(\mu, {\bm\sigma^2}, {\bm T}, {\bm\theta})$ with respect to $\mu$ to be zero, we have the maximum likelihood estimator of $\mu$ is
	\begin{align}\label{mu}
		\hat{\mu}=\frac{{\bm 1}_n^T{\bm\Phi}^{-1}{\bm y}_n}{{\bm 1}_n^T{\bm\Phi}^{-1}{\bm 1}_n}.
	\end{align}
	Substituting (\ref{mu}) into (\ref{loglikelihood}), we obtain
	\begin{align*}
		l(\hat\mu, {\bm\sigma^2}, {\bm T}, {\bm\theta})
		&=-\frac n2\log(2\pi)-\frac 12\log|{\bm\Phi}|-\frac 12({\bm y}_n-\hat\mu{\bm 1}_n)^T{\bm\Phi}^{-1}({\bm y}_n-\hat\mu{\bm 1}_n)\\
		&=-\frac n2\log(2\pi)-\frac 12\log|{\bm\Phi}|-\frac 12 {\bm y}_n^{T}{\bm\Phi}^{-1}{\bm y}_n+\frac 12\frac{({\bm 1}_n^T{\bm\Phi}^{-1}{\bm y}_n)^2}{{\bm 1}_n^T{\bm\Phi}^{-1}{\bm 1}_n}.
	\end{align*}
	The estimators of ${\bm\sigma^2}, {\bm T}, {\bm\theta}$ can be obtained as
	\begin{align}\label{mini}
		(\hat{\bm\sigma}^2, \hat{\bm T}, \hat{\bm\theta})=\argmin\left\{\log|{\bm\Phi}|+{\bm y}_n^{T}{\bm\Phi}^{-1}{\bm y}_n-\frac{({\bm 1}_n^T{\bm\Phi}^{-1}{\bm y}_n)^2}{{\bm 1}_n^T{\bm\Phi}^{-1}{\bm 1}_n}\right\}.
	\end{align}
	
	To ensure the $m_j\times m_j$ matrix ${\bm T}^{(j)}$ is a valid correlation function, ${\bm T}^{(j)}=(\tau_{r,s}^{(j)})_{m_j\times m_j}$ must be a positive definite matrix with unit diagonal elements. We apply the hypersphere parameterization approach in Zhou et al. (2011) to quantify the correlations of qualitative factors. The key of this approach is to find a lower triangular matrix ${\bm L}^{(j)}=(l_{r,s}^{(j)})$ with strictly positive diagonal entries such that
	\begin{align*}
		{\bm T}^{(j)}={\bm L}^{(j)}{\bm L}^{(j)T}.
	\end{align*}
	For $r=1$, let $l_{1,1}=1$. For $r=2,\cdots, m_j$, the entries of the $r$th row of ${\bm L}^{(j)}$ can be constructed as follows:
	\begin{align*}
		&l_{r,1}^{(j)}=\cos(\theta_{r,1}),\\
		&l_{r,2}^{(j)}=\sin(\theta_{r,1})\cos(\theta_{r,2}),\\
		&\cdots,\\
		&l_{r,r-1}^{(j)}=\sin(\theta_{r,1})\cdots\sin(\theta_{r,r-2})\cos(\theta_{r,r-1}),\\
		&l_{r,r}^{(j)}=\sin(\theta_{r,1})\cdots\sin(\theta_{r,r-2})\sin(\theta_{r,r-1}),
	\end{align*}
	where $\theta_{r,s}\in(0,\pi)$.
	
	\vspace{2ex}
	\noindent{\bf Proof of Lemma \ref{prop2}}
	\begin{proof}

Recall the definition of $y_{\min}, \tilde\mu_{\min,n},  \tilde\mu_{\min,n}^L, \tilde\mu_{\min,n}^U$ as follows
\begin{align}
		&y_{\min}=\min_{\bm w_0\in\mathcal{A}}y({\bm w}_0)=\sup_{v\in R}\{v: \frac{1}{M}\sum_{{\bm z}_0\in {\bm Z}}\int_{{\bm x}_0\in {\bm X}}{\bm 1}_{\{y({\bm w}_0)< v\}} {\rm d}{\bm x}_0<\epsilon\},\label{esti1}\\
		&\tilde\mu_{\min,n}=\min_{\bm w_0\in\mathcal{A}}\mu_{0|n}({\bm w}_0)=\sup_{v\in R}\{v: \frac{1}{M}\sum_{{\bm z}_0\in {\bm Z}}\int_{{\bm x}_0\in {\bm X}}{\bm 1}_{\{\mu_{0|n}({\bm w}_0)< v\}}{\rm d}{\bm x}_0<\epsilon\},\label{esti2}\\
		&\tilde\mu_{\min,n}^L=\min_{\bm w_0\in\mathcal{A}}\mu_{0|n}^L({\bm w}_0)=\sup_{v\in R}\{v: \frac{1}{M}\sum_{{\bm z}_0\in {\bm Z}}\int_{{\bm x}_0\in {\bm X}}{\bm 1}_{\{\mu_{0|n}^L({\bm w}_0)<v\}}{\rm d}{\bm x}_0<\epsilon\},\label{esti3}\\
		&\tilde\mu_{\min,n}^U=\min_{\bm w_0\in\mathcal{A}}\mu_{0|n}^U({\bm w}_0)=\sup_{v\in R}\{v: \frac{1}{M}\sum_{{\bm z}_0\in {\bm Z}}\int_{{\bm x}_0\in {\bm X}}{\bm 1}_{\{\mu_{0|n}^U({\bm w}_0)< v\}}{\rm d}{\bm x}_0<\epsilon\},\label{esti4}
	\end{align}
where $\epsilon>0$ is a small positive number, and ${\bm 1}_{\{\cdot\}}$ is the indicator function.

Let $\alpha'=\alpha/2$. By Lemma \ref{prop1},  for a given ${\bm x}_0\in{\bm X}$, we have
		\begin{align*}
			P\left(\mu_{0|n}^L({\bm w}_0)\leq y({\bm w}_0)\leq \mu_{0|n}^U({\bm w}_0),\ \ \forall {\bm z}_0\in {\bm Z}, \forall n \geq 1\right)\geq 1-\alpha'.
		\end{align*}
		When $\mu_{0|n}^L({\bm w}_0)\leq y({\bm w}_0)\leq \mu_{0|n}^U({\bm w}_0)$, for given  ${\bm x}_0\in{\bm X}$ and all $v\in R$,
		\begin{align*}
			\frac{1}{M}\sum_{{\bm z}_0\in {\bm Z}}{\bm 1}_{\{\mu_{0|n}^U({\bm w}_0)< v\}}\leq\frac{1}{M}\sum_{{\bm z}_0\in {\bm Z}}{\bm 1}_{\{y({\bm w}_0)< v\}}\leq\frac{1}{M}\sum_{{\bm z}_0\in {\bm Z}}{\bm 1}_{\{\mu_{0|n}^L({\bm w}_0)< v\}},\forall n \geq 1.
		\end{align*}
		Integrating each term over  ${\bm x}_0$ in ${\bm X}$, we have
		\begin{align}\label{v}
			\int_{{\bm x}_0\in {\bm X}}\frac{1}{M}\sum_{{\bm z}_0\in {\bm Z}}{\bm 1}_{\{\mu_{0|n}^U({\bm w}_0)< v\}}{\rm d}{\bm x}_0&\leq\int_{{\bm x}_0\in {\bm X}}\frac{1}{M}\sum_{{\bm z}_0\in {\bm Z}}{\bm 1}_{\{y({\bm w}_0)< v\}}{\rm d}{\bm x}_0\notag\\
			&\leq\int_{{\bm x}_0\in {\bm X}}\frac{1}{M}\sum_{{\bm z}_0\in {\bm Z}}{\bm 1}_{\{\mu_{0|n}^L({\bm w}_0)< v\}}{\rm d}{\bm x}_0,\forall n \geq 1.
		\end{align}
		
		Let $v=\tilde\mu_{\min,n}^L$ given in (\ref{esti3}), we get
		\begin{align*}
			\int_{{\bm x}_0\in {\bm X}}\frac{1}{M}\sum_{{\bm z}_0\in {\bm Z}}{\bm 1}_{\{y({\bm w}_0)< \tilde\mu_{\min,n}^L\}}{\rm d}{\bm x}_0<\epsilon.
		\end{align*}
		By \eqref{esti1}, we obtain that $\tilde\mu_{\min,n}^L\leq y_{\min}.$ Thus
		\begin{align}
			&P\left(\tilde\mu_{\min,n}^L\leq y_{\min}, \forall n \geq 1\right)\notag\\
			\geq &P\left(\mu_{0|n}^L({\bm w}_0)\leq y({\bm w}_0)\leq \mu_{0|n}^U({\bm w}_0),\forall {\bm z}_0\in {\bm Z}, \forall n \geq 1\right)\notag\\
			\geq &1-\alpha'.\label{prop21}
		\end{align}
		
		Similarly, in (\ref{v}), let $v= y_{\min}$, we get that
		\begin{align*}
			\int_{{\bm x}_0\in {\bm X}}\frac{1}{M}\sum_{{\bm z}_0\in {\bm Z}}{\bm 1}_{\{\mu_{0|n}^U({\bm w}_0)< y_{\min}\}}{\rm d}{\bm x}_0<\epsilon.
		\end{align*}
		By (\ref{esti4}), we obtain that $y_{\min}\leq\tilde\mu_{\min,n}^U,$ thus
		\begin{align}
			&P\left(y_{\min}\leq\tilde\mu_{\min,n}^U, \forall n \geq 1\right)\notag\\
			\geq &P\left(\mu_{0|n}^L({\bm w}_0)\leq y({\bm w}_0)\leq \mu_{0|n}^U({\bm w}_0),\forall {\bm z}_0\in {\bm Z}, \forall n \geq 1\right)\notag\\
			\geq &1-\alpha'.\label{prop22}
		\end{align}
		By \eqref{prop21} and \eqref{prop22}, we have
		\begin{align*}
			&P\left(y_{\min}\in[\tilde\mu_{\min,n}^L, \tilde\mu_{\min,n}^U],\forall n\geq 1\right)\\
			= &P\left(\tilde\mu_{\min,n}^L\leq y_{\min}\leq\tilde\mu_{\min,n}^U,\forall n\geq 1\right)\\
			\geq &1-2\alpha'=1-\alpha.
		\end{align*}
	\end{proof}
	
	\vspace{2ex}
	\noindent{\bf Proof of Theorem \ref{prop3}}
	\begin{proof}
		By (\ref{esti4}), for any $n \geq 1$,
		\begin{align*}
			\frac{1}{M}\sum_{{\bm z}_0\in {\bm Z}}\int_{{\bm x}_0\in {\bm X}}{\bm 1}_{\{\mu_{0|n}^U({\bm w}_0)< \tilde\mu_{\min,n}^U\}}{\rm d}{\bm x}_0<\epsilon.
		\end{align*}
		Define
		$$\mathcal{A}_{n}(\bm z_0)=\{{\bm x}_0\in \bm X: \mu_{0|n}^L({\bm x}_0,\bm z_0)\leq \tilde\mu_{\min,n}^U\}.$$
		It is clear that $\mathcal{A}_{n}=\{(\bm x_0, \bm z_0): \bm z_0\in \bm Z, \bm x_0\in \mathcal{A}_{n}(\bm z_0)\}$, then we have
		\begin{align*}
			\frac{1}{M}\sum_{{\bm z}_0\in {\bm Z}}\int_{{\bm x}_0\in \mathcal{A}_{n}(\bm z_0)}{\bm 1}_{\{\mu_{0|n}^U({\bm w}_0)< \tilde\mu_{\min,n}^U\}}{\rm d}{\bm x}_0=\frac{1}{M}\sum_{{\bm z}_0\in {\bm Z}}\int_{{\bm x}_0\in {\bm X}}{\bm 1}_{\{\mu_{0|n}^U({\bm w}_0)< \tilde\mu_{\min,n}^U\}}{\rm d}{\bm x}_0<\epsilon.
		\end{align*}
		Since $\mu_{0|n}^U({\bm w}_0)=\mu_{0|n}({\bm w}_0)+\sqrt{\beta_{0|n}}\sigma_{0|n}({\bm w}_0)$, we get that
		\begin{align*}
			&\frac{1}{M}\sum_{{\bm z}_0\in {\bm Z}}\int_{{\bm x}_0\in \mathcal{A}_{n}(\bm z_0)}{\bm 1}_{\{\mu_{0|n}({\bm w}_0)< \tilde\mu_{\min,n}^U-\sqrt{\beta_{0|n}}\sup_{{\bm w}_0\in \mathcal{A}_{n}}\sigma_{0|n}({\bm w}_0)\}}{\rm d}{\bm x}_0\\
			\leq &\frac{1}{M}\sum_{{\bm z}_0\in {\bm Z}}\int_{{\bm x}_0\in \mathcal{A}_{n}(\bm z_0)}{\bm 1}_{\{\mu_{0|n}({\bm w}_0)< \tilde\mu_{\min,n}^U-\sqrt{\beta_{0|n}}\sigma_{0|n}({\bm w}_0)\}}{\rm d}{\bm x}_0\\
			<&\epsilon.
		\end{align*}
		By the definition of $\mathcal{A}_{n}(\bm z_0)$,
		\begin{align*}
			&\frac{1}{M}\sum_{{\bm z}_0\in {\bm Z}}\int_{{\bm x}_0\in {\bm X}}{\bm 1}_{\{\mu_{0|n}({\bm w}_0)< \tilde\mu_{\min,n}^U-\sqrt{\beta_{0|n}}\sup_{{\bm w}_0\in \mathcal{A}_{n}}\sigma_{0|n}({\bm w}_0)\}}{\rm d}{\bm x}_0\\
			=&\frac{1}{M}\sum_{{\bm z}_0\in {\bm Z}}\int_{{\bm x}_0\in \mathcal{A}_{n}(\bm z_0)}{\bm 1}_{\{\mu_{0|n}({\bm w}_0)< \tilde\mu_{\min,n}^U-\sqrt{\beta_{0|n}}\sup_{{\bm w}_0\in \mathcal{A}_{n}}\sigma_{0|n}({\bm w}_0)\}}{\rm d}{\bm x}_0\\
			<&\epsilon.
		\end{align*}
		Thus,
		\begin{align*}
			\frac{1}{M}\sum_{{\bm z}_0\in {\bm Z}}\int_{{\bm x}_0\in {\bm X}}{\bm 1}_{\{\mu_{0|n}({\bm w}_0)< \tilde\mu_{\min,n}^U-\sqrt{\beta_{0|n}}\sup_{{\bm w}_0\in \mathcal{A}_{n}}\sigma_{0|n}({\bm w}_0)\}}{\rm d}{\bm x}_0<\epsilon.
		\end{align*}
		By (\ref{esti2}), we obtain that
		\begin{align*}
			\tilde\mu_{\min,n}\geq \tilde\mu_{\min,n}^U-\sqrt{\beta_{0|n}}\sup_{{\bm w}_0\in \mathcal{A}_{n}}\sigma_{0|n}({\bm w}_0).
		\end{align*}
		Let $\alpha'=\alpha/2$, by Lemma \ref{prop2}, we have $P\left(\tilde\mu_{\min,n}^U\geq y_{\min},\forall n\geq 1\right)\geq 1-\alpha'.$ Thus,
		\begin{align}\label{prop3fo1}
			P\left(\tilde\mu_{\min,n}\geq  y_{\min}-\sqrt{\beta_{0|n}}\sup_{{\bm w}_0\in \mathcal{A}_{n}}\sigma_{0|n}({\bm w}_0), \forall n\geq 1\right)\geq 1-\alpha',
		\end{align}
		where  $\beta_{0|n}=2\log\frac{\pi^2n^2M}{6\alpha'}$.
		
		On the other hand, by (\ref{esti2}), for any $n \geq 1$,
		\begin{align*}
			\frac{1}{M}\sum_{{\bm z}_0\in {\bm Z}}\int_{{\bm x}_0\in {\bm X}}{\bm 1}_{\{\mu_{0|n}({\bm w}_0)< \tilde\mu_{\min,n}\}}{\rm d}{\bm x}_0<\epsilon.
		\end{align*}
		By the definition of $\mathcal{A}_{n}(\bm z_0)$ and $\tilde \mu ^\mathcal{A}_{n}\leq\tilde\mu_{\min,n}\leq\tilde \mu ^U_{\min,n}$, we get that
		\begin{align*}
			\frac{1}{M}\sum_{{\bm z}_0\in {\bm Z}}\int_{{\bm x}_0\in \mathcal{A}_{n}(\bm z_0)}{\bm 1}_{\{\mu_{0|n}({\bm w}_0)< \tilde\mu_{\min,n}\}} {\rm d}{\bm x}_0=\frac{1}{M}\sum_{{\bm z}_0\in {\bm Z}}\int_{{\bm x}_0\in {\bm X}}{\bm 1}_{\{\mu_{0|n}({\bm w}_0)< \tilde\mu_{\min,n}\}}{\rm d}{\bm x}_0<\epsilon.
		\end{align*}
		Since $\mu_{0|n}^L({\bm w}_0)=\mu_{0|n}({\bm w}_0)-\sqrt{\beta_{0|n}}\sigma_{0|n}({\bm w}_0)$, we get that
		\begin{align*}
			&\frac{1}{M}\sum_{{\bm z}_0\in {\bm Z}}\int_{{\bm x}_0\in \mathcal{A}_{n}(\bm z_0)}{\bm 1}_{\{\mu_{0|n}^L({\bm w}_0) < \tilde\mu_{\min,n}-\sqrt{\beta_{0|n}}\sup_{{\bm w}_0\in \mathcal{A}_{n}}\sigma_{0|n}({\bm w}_0)\}}{\rm d}{\bm x}_0\\
			=&\frac{1}{M}\sum_{{\bm z}_0\in {\bm Z}}\int_{{\bm x}_0\in \mathcal{A}_{n}(\bm z_0)}{\bm 1}_{\{\mu_{0|n}({\bm w}_0)-\sqrt{\beta_{0|n}}\sigma_{0|n}({\bm w}_0) < \tilde\mu_{\min,n}-\sqrt{\beta_{0|n}}\sup_{{\bm w}_0\in \mathcal{A}_{n}}\sigma_{0|n}({\bm w}_0)\}}{\rm d}{\bm x}_0\\
			\leq&\frac{1}{M}\sum_{{\bm z}_0\in {\bm Z}}\int_{{\bm x}_0\in \mathcal{A}_{n}(\bm z_0)}{\bm 1}_{\{\mu_{0|n}({\bm w}_0)-\sqrt{\beta_{0|n}}\sup_{{\bm w}_0\in \mathcal{A}_{n}}\sigma_{0|n}({\bm w}_0) < \tilde\mu_{\min,n}-\sqrt{\beta_{0|n}}\sup_{{\bm w}_0\in \mathcal{A}_{n}}\sigma_{0|n}({\bm w}_0)\}}{\rm d}{\bm x}_0\\
			<&\epsilon.
		\end{align*}
		By $\tilde \mu ^\mathcal{A}_{n}\leq\tilde\mu_{\min,n}\leq\tilde \mu ^U_{\min,n}$ and the definition of $\mathcal{A}_{n}(\bm z_0)$,
		\begin{align*}
			&\frac{1}{M}\sum_{{\bm z}_0\in {\bm Z}}\int_{{\bm x}_0\in \bm X}{\bm 1}_{\{\mu_{0|n}^L({\bm w}_0)< \tilde\mu_{\min,n}-\sqrt{\beta_{0|n}}\sup_{{\bm w}_0\in \mathcal{A}_{n}}\sigma_{0|n}({\bm w}_0)\}}{\rm d}{\bm x}_0\\
			=&\frac{1}{M}\sum_{{\bm z}_0\in {\bm Z}}\int_{{\bm x}_0\in \mathcal{A}_{n}(\bm z_0)}{\bm 1}_{\{\mu_{0|n}^L({\bm w}_0)< \tilde\mu_{\min,n}-\sqrt{\beta_{0|n}}\sup_{{\bm w}_0\in \mathcal{A}_{n}}\sigma_{0|n}({\bm w}_0)\}}{\rm d}{\bm x}_0\\
			<&\epsilon.
		\end{align*}
		Thus,
		\begin{align*}
			\frac{1}{M}\sum_{{\bm z}_0\in {\bm Z}}\int_{{\bm x}_0\in \bm X}{\bm 1}_{\{\mu_{0|n}^L({\bm w}_0)< y_{\min}-\sqrt{\beta_{0|n}}\sup_{{\bm w}_0\in \mathcal{A}_{n}}\sigma_{0|n}({\bm w}_0)\}}{\rm d}{\bm x}_0
			<\epsilon.
		\end{align*}
		By (\ref{esti3}), we obtain that
		\begin{align*}
			\tilde\mu_{\min,n}^L\geq \tilde\mu_{\min,n}-\sqrt{\beta_{0|n}}\sup_{{\bm w}_0\in \mathcal{A}_{n}}\sigma_{0|n}({\bm w}_0).
		\end{align*}
		By Lemma \ref{prop2}, we have
		\begin{align}\label{prop3fo2}
			P\left(y_{\min}\geq \tilde\mu_{\min,n}-\sqrt{\beta_{0|n}}\sup_{{\bm w}_0\in \mathcal{A}_{n}}\sigma_{0|n}({\bm w}_0),\forall n\geq 1\right)\geq 1-\alpha'.
		\end{align}
		Using (\ref{prop3fo1}) and (\ref{prop3fo2}), the conclusion of Theorem \ref{prop3} is proved.
	\end{proof}
	
	\vspace{2ex}
	\noindent{\bf Proof of Corollary \ref{prop4}}
	
	\begin{proof}
		Let $\alpha'=\alpha/2$, by Lemma \ref{prop1},  for the minimum point ${\bm w}^*\in\mathcal{A}$, we have
		\begin{align*}
			P\left(\mu_{0|n}^L({\bm w}^*)\leq y({\bm w}^*)\leq \mu_{0|n}^U({\bm w}^*),\forall n \geq 1\right)\geq 1-\alpha'.
		\end{align*}
		By the definition of $ y_{\min}$ in \eqref{esti1}, we have $y_{\min}=y({\bm w}^*)$. Thus
		$$P\left(\mu_{0|n}^L({\bm w}^*)\leq y_{\min}\leq \mu_{0|n}^U({\bm w}^*),\forall n \geq 1\right)\geq 1-\alpha'.$$
		
		By Lemma \ref{prop2}, we have
		\begin{align*}
			P\left(\tilde\mu_{\min,n}^L\leq y_{\min}\leq\tilde\mu_{\min,n}^U,\forall n\geq 1\right)\geq 1-\alpha'.
		\end{align*}
		Thus
		\begin{align*}
			&P\left(\mu_{0|n}^L({\bm w}^*)\leq\tilde\mu_{\min,n}^U,\forall n\geq 1\right)\\
			\geq &P\left(\mu_{0|n}^L({\bm w}^*)\leq y_{\min}\leq \mu_{0|n}^U({\bm w}^*),\tilde\mu_{\min,n}^L\leq y_{\min}\leq\tilde\mu_{\min,n}^U,\forall n\geq 1\right)\\
			\geq &1-2\alpha'=1-\alpha.
		\end{align*}
		It implies the conclusion of Corollary \ref{prop4}.
	\end{proof}
	
	\bibliographystyle{apalike}

\begin{thebibliography}{}	
		
		\bibitem[Bichon et al, 2009]{Bi09}
		Bichon, B., Mahadevan, S., and Eldred, M. (2009).
		\newblock Reliability-based design optimization using efficient global reliability analysis.
		\newblock {\em  50th AIAA/ASME/ASCE/AHS/ASC structures, structural dynamics, and materials conference, AIAA, Palm Springs, California}, 2009-2261.
		
		\bibitem[Bingham et al., 2014)]{Bi14}
		Bingham, D., Ranjan, P., and Welch, W. J. (2014).
		\newblock Design of computer experiments for optimization, estimation of function contours, and related objectives.
		\newblock {\em  Statistics in Action: A Canadian Outlook}, 109.
		
		\bibitem[Cameron et al, 2019]{Ca19}
		Cameron, K.W., Anwar, A., Cheng, Y., et al. (2019).
		\newblock MOANA: modeling and analyzing I/O variability in parallel system experimental design.
		\newblock {\em IEEE Transactions on Parallel and Distributed Systems}, 30(8), 1843-1856.
		
		\bibitem[Chen et al, 2019]{Ch19}
		Chen, Z., Mak, S., and Wu, C. F. J. (2019)
		\newblock   A hierarchical expected improvement method for Bayesian optimization.
		\newblock {\em  arXiv preprint arXiv}: 1911.07285.
		
		
		\bibitem[Deng et al, 2015]{De15}
		Deng, X., Hung, Y. and Lin, C. D. (2015).
		\newblock Design for computer experiments with qualitative and quantitative factors.
		\newblock {\em  Statistica Sinica}, 25(4), 1567-1581.
		
		\bibitem[Deng et al, 2017]{De17}
		Deng, X., Lin, C. D., Liu, K. W., and Rowe, R. K. (2017).
		\newblock Additive Gaussian process for computer models with qualitative and quantitative factors.
		\newblock {\em  Technometrics}, 59, 283-292.
		
		\bibitem[Fang et al, 2005]{Fa05}
		Fang, K. T., Li, R., and Sudjianto, A. (2005).
		\newblock {\em Design and Modeling for Computer Experiments},
		\newblock  New York: Chapman \& Hall/CRC Press.
		
		\bibitem[Frazier et al, 2017]{Fr17}
		Frazier, P. I., Powell, W. B., and Dayanik, S. (2008). 
		\newblock A knowledge-gradient policy for sequential information collection. 
		\newblock {\em SIAM Journal on Control and Optimization}, 47(5), 2410-2439.
		
        \bibitem[Gramacy, 2020]{Gr20}
		 Gramacy, R. B. (2020). 
		 \newblock {\em Surrogates: Gaussian Process Modeling, Design, and Optimization for the Applied Sciences},
		 \newblock New York: Chapman \& Hall/CRC Press.
		
		\bibitem[He et al, 2017]{He17}
		He, Y., Lin, C. D., Sun, F., and Lv, B. (2017).
		\newblock Marginally coupled designs for two-level qualitative factors.
		\newblock {\em  Journal of  Statistical Planning and Inference}, 187, 103-108.
		
		\bibitem[He et al, 2019]{He19}
		He, Y., Lin, C.D., Sun, F., and Lv, B. (2019).
		\newblock Construction of marginally coupled designs by subspace theory.
		\newblock {\em  Bernoulli},   25, 2163-2182.
		
		\bibitem[Jala et al, 2016]{Ja16}
		Jala, M., Levy-Leduc, C., Moulines, E., Conil, E., and Wiart, J. (2016).
		\newblock Sequential design of computer experiments for the assessment of fetus exposure to electromagnetic fields.
		\newblock {\em  Technometrics},  58, 30-42.
		
		\bibitem[Joseph, 2016]{Jo16}
		Joseph, V. R. (2016).
		\newblock Space-filling designs for computer experiments: A review.
		\newblock {\em  Quality Engineering},  28(1), 28-35.

		
		\bibitem[Jone et al, 1998]{J098}
		Jones, D. R., Schonlau, M., and Welch, W. J. (1998).
		\newblock Efficient global optimization of expensive black-box functions.
		\newblock {\em  Journal of Global Optimization},  13, 455-492.

		\bibitem[Lin and Tang, 2015]{Li15}
		Lin, C. D., and Tang, B. (2015).
		\newblock Latin hypercubes and space-filling designs.
		\newblock {\em  Handbook of Design and Analysis of Experiments},  593-625, Chapman \& Hall/CRC, London.
		
		\bibitem[McKay et al, 1979]{Mc79}
		McKay, M. D., Beckman, R. J., and Conover, W. J. (1979).
		\newblock A comparison of three methods for selecting values of input variables in the analysis
		of output from a computer code.
		\newblock {\em  Technometrics}, 21, 239-245.
		
		\bibitem[Norcott, W. D., 2020]{Norcott20}
		Norcott, W. D. (2020).
		\newblock IOzone filesystem benchmark.
		\newblock {\href{http://www.iozone.org/}{http://www.iozone.org/}}.
		
		\bibitem[Ponweiser, 2016]{Po08}
		Ponweiser, W., Wagner, T., and Vincze, M. (2008).
		\newblock    Clustered multiple generalized expected improvement: A novel infill sampling criterion for surrogate models.
		\newblock {\em  2008 IEEE Congress on Evolutionary Computation}: 3515-3522.
		
		\bibitem[Picheny, 2016]{Pi16}
		Picheny, V., Gramacy, R.B., Wild, S.M., and Digabel S.L. (2016).
		\newblock   Bayesian optimization under mixed constraints with a slack-variable augmented Lagrangian.
		\newblock {\em  arXiv preprint arXiv}: 1605.09466.
		
		\bibitem[Qian, 2012]{Qi12}
		Qian, P. Z. (2012).
		\newblock Sliced Latin hypercube designs.
		\newblock {\em  Journal of the American Statistical Association}, 107(497), 393-399.
		
		\bibitem[Ranjan et al., 2008]{Ra08}
		Ranjan, P., Bingham, D., and Michailidis, G. (2008).
		\newblock Sequential experiment design for contour estimation from complex computer codes.
		\newblock {\em  Technometrics}, 50(4), 527-541.
		
		\bibitem[Ranjan et al., 2011]{Ra11}
		Ranjan, P., Haynes, R., and Karsten, R. (2011).
		\newblock A computationally stable approach to Gaussian process interpolation of deterministic computer simulation data.
		\newblock {\em  Technometrics}, 53(4), 366-378.
		
		\bibitem[Roy, 2008]{Ro08}
		Roy, S. (2008).
		\newblock {\em Sequential-adaptive design of computer experiments for the estimation of percentiles}.
		\newblock Ohio State University.  Ph.D. Thesis.
		
		\bibitem[Sacks et al., 1989]{Sa89}
		Sacks, J., Welch, W. J., Mitchell, T. J., and Wynn., H. P. (1989).
		\newblock Design and analysis of computer experiments.
		\newblock {\em  Statistical Science}, 4, 409-423.
		
		\bibitem[Santner et al., 2003]{Sa03}
		Santner, T. J., Williams, B. J., and Notz,W. I. (2003).
		\newblock {\em The Design and Analysis of Computer Experiments.}
		\newblock New York: Springer.
		
		\bibitem[Sakellariou et al., 2018]{Sa18}
		Sakellariou, R., Buenabad-Ch\'{a}vez, J., Kavakli, E., Spais, I., and Tountopoulos, V. (2018).
		\newblock High performance computing and industry 4.0: experiences from the {DISRUPT} project.
		\newblock {\em  SAMOS '18: Proceedings of the 18th International Conference on Embedded Computer Systems: Architectures, Modeling, and Simulation}, Pages 218–219, DOI: 10.1145/3229631.3264660.
		
		\bibitem[Sauer et al., 2020]{Sa20}
		Sauer A., Gramacy R. B., and Higdon D. (2020).
		\newblock Active learning for deep Gaussian process surrogates.
		\newblock {\em  arXiv preprint arXiv}: 2012.08015.
		
		\bibitem[Schonlau et al., 1998]{Sc98}
		Schonlau, M., Welch, W. J., and Jones, D. R. (1998).
		\newblock New developments and applications in experimental design.
		\newblock {\em  Global versus local search in constrained optimization of computer models, Lecture Notes-Monograph Series}: 34, 11-25.
		
		\bibitem[Scott et al., 2011]{Sc11}
		Scott, W., Frazier, P., and Powell, W. (2011). 
		\newblock The correlated knowledge gradient for simulation optimization of continuous parameters using Gaussian process regression. 
		\newblock {\em SIAM Journal on Optimization}, 21(3), 996-1026.
		
		\bibitem[Sobester et al., 2005]{So05}
		Sobester, A., Leary, S. J., and Keane, A. J. (2005).
		\newblock On the design of optimization strategies based on global response surface approximation models.
		\newblock {\em  Journal of Global Optimization}: 33(1), 31–59.
		
		\bibitem[Srinivas et al., 2010]{Sr10}
		Srinivas, N., Krause, A., Kakade, S. M., and Seeger, M. (2010).
		\newblock Gaussian process optimization in the bandit setting: no regret and experimental design.
		\newblock {\em  arXiv preprint arXiv}: 0912.3995.
			
		\bibitem[Sung et al, 2020]{Sung20}
		Sung, C. L., Hung, Y., Rittase, W., Zhu, C., and Wu, C. F. J. (2020). 
		\newblock A generalized Gaussian process model for computer experiments with binary time series. 
		\newblock {\em Journal of the American Statistical Association}, 115(530), 945-956.
		
		\bibitem[Wang et al, 2018]{Wa18}
		Wang, L., Xiao, Q., and Xu, H. (2018).
		\newblock Optimal maximin $L_1$ distance Latin hypercube designs based on good lattice point designs.
		\newblock {\em  The Annals of Statistics}, 46, 3741-3766.

		\bibitem[Wu and Hamada, 2009]{Wu09}
		Wu, C. F. J., and Hamada, M. (2009).
		\newblock {\em Experiments: Planning, Analysis, and Optimization}.
		\newblock New York: Wiley.
		
		\bibitem[Xiao and Xu, 2017]{Xi17}
		Xiao, Q., and Xu, H. (2017).
		\newblock Construction of maximin distance Latin squares and related Latin hypercube designs.
		\newblock {\em  Biometrika}, 104, 455-464.	
		

		\bibitem[Xie and Deng, 2020]{Xie20}
		 Xie, W., and Deng, X. (2020). 
		 \newblock Scalable algorithms for the sparse ridge regression.
		 \newblock {\em SIAM Journal on Optimization}, 30(4), 3359-3386.

		\bibitem[Xu et al, 1979]{Xu20}
		Xu, L., Lux, T., Chang, T., Li, B., Hong, Y., Watson, L., Butt, A., Yao, D., and Cameron, K. (2020).
		\newblock Prediction of high-performance computing input/output variability and its application to optimization for system configurations.
		\newblock {\em  Quality Engineering}, DOI:10.1080/08982112.2020.1866203.

		\bibitem[Yang et al., 2020]{Ya20}
		Yang, F., Lin, C. D., and Ranjan, P. (2020).
		\newblock Global fitting of the response surface via estimating multiple contours of a simulator.
		\newblock {\em  Journal of Statistical Theory and Practice}, 14, 1-21.
		
		\bibitem[Zhang et al., 2020]{Zh20}
		Zhang, Q., Chien, P., Liu Q., Xu L., and Hong, Y. (2020).
		\newblock Mixed-input Gaussian process emulators for computer experiments with a large number of categorical levels.
		\newblock {\em  Journal of Quality Technology}, DOI: 10.1080/00224065.2020.1778431.
		
		\bibitem[Zhou et al., 2011]{Zh11}
		Zhou, Q., Qian, P .Z. G., and Zhou, S. (2011).
		\newblock A simple approach to emulation for computer models with qualitative and quantitative factors.
		\newblock {\em  Technometrics}, 53, 266-273.
	\end{thebibliography}

\end{document}